\documentclass[11pt]{article}

\usepackage{amsfonts}
\usepackage{amssymb}
\usepackage{amsmath}
\usepackage{authblk}
\usepackage{bbm}
\usepackage{hyperref}
\usepackage{graphicx}
\usepackage{float}
\usepackage{overpic}
\usepackage{geometry}
\usepackage{epstopdf}
\usepackage{multirow}
\usepackage{subcaption}
\usepackage[official]{eurosym}
\providecommand{\U}[1]{\protect\rule{.1in}{.1in}}
\usepackage{caption}
\usepackage{subcaption}

\newtheorem{theorem}{Theorem}[section]

\newtheorem{lemma}[theorem]{Lemma}

\newtheorem{proposition}[theorem]{Proposition}
\newtheorem{remark}[theorem]{Remark}

\newenvironment{proof}[1][Proof]{\noindent\textbf{#1.} }{\ \rule{0.5em}{0.5em}}
\newcommand{\R}{{\mathbb R}}
\newcommand{\tr}{\mathrm{tr}}

\usepackage{amsmath,systeme}
\usepackage{lineno}
\usepackage[dvipsnames]{xcolor}
\usepackage{indentfirst}

\title{Optimal energy storage management for self-consumption groups}
\author{Awerkin, A.\footnote{Department of Economics and Management, University of Pavia, {\tt almendraalexandra.awerkinvargas@unipv.it}} \footnote{Corresponding author.}, De Giuli, M.E.\footnote{Department of Economics and Management, University of Pavia, {\tt elena.degiuli@unipv.it}}, Vargiolu T.\footnote{Department of Mathematics ``Tullio Levi Civita”, University of Padova, {\tt vargiolu@math.unipd.it}}}

\begin{document}
\maketitle

\begin{abstract}
We study the optimal management of a photovoltaic system's battery owned by a self-consumption group that aims to minimize energy consumption costs. We assume that the photovoltaic system is composed of a photovoltaic panel and a battery, where the photovoltaic panel produces energy according to a certain stochastic process. The management of the  battery is the responsibility of a group administrator, who makes the joint decision to either store part of the photovoltaic energy production and sell the remaining energy at the electricity spot price, or discharge part of the energy stored in the battery and sell it in the electricity market. Inspired by European Union and Italian legislation, which promote incentives for energy transition and renewable energy production, we assume that the group receives a monetary incentive for the virtual self-consumed energy, defined as the minimum between the power bought from the grid to satisfy the group's power demand and the energy sold to the market. In this case, the energy sold by the group is a mix of part of the photovoltaic production that is not stored and part of the energy discharged from the battery. We model the problem as a stochastic optimal control problem, where the optimal strategy is the joint charge-discharge decision that minimizes the group's energy consumption costs. We find the solution numerically by applying a finite difference scheme to solve the Hamilton-Jacobi-Bellman equation associated with the value  function of the optimal control problem. 
\end{abstract}

\noindent\textbf{Keywords}\\ Self-consumption group, optimal stochastic control, photovoltaic system, photovoltaic modeling.

\section{Introduction}
 
The self-production and self-consumption of renewable energy is becoming pivotal in the transition towards a more sustainable and decentralized energy system. It leverages local renewable energy sources and storage systems to enhance energy self-sufficiency and reduce greenhouse gas emissions, consistent with the goals of the Paris Agreement, which aims to achieve net-zero climate impact by 2050.
By reducing reliance on fossil fuels and decreasing carbon footprints, self-consumption groups and renewable energy communities (REC) contribute to national and international efforts to meet these climate targets. Moreover, their localized nature fosters community engagement and resilience, further promoting the extensive implementation of renewable energy solutions essential for achieving the long-term goals of the Paris Agreement. In this line, in 2019 the European Union adopted the Clean Energy Package for all Europeans, which consists in 8 laws, establishing the directives and legal definition in order to regularize and promote the creation of self-consumption groups and RECs. Hence, self-consumption groups and RECs are driven by the EU's regulatory framework. 

According to the EU Directive 2018/2001 (REDII) \cite{EU2018/2001}, self-consumption groups consist of individuals or entities that produce renewable energy, with the primary goal of reducing energy costs and increasing the energy self-sufficiency of group members. Typically, the members of a self-consumption group are located close to each other and share energy production infrastructures, such as solar panels installed on residential or commercial buildings. While RECs involve various stakeholders in a collaborative effort to produce, manage, and consume renewable energy, self-consumption groups primarily focus on individual production and consumption.

Moreover, the EU Directive 2018/2001 (RED II)\cite{EU2018/2001}, promotes self-consumption and renewable energy communities by defining the rights to produce, store, and sell renewable energy. On the other hand, the EU Directive 2019/944 (IEMD) \cite{2019/944}, introduces the concept of the "active citizen", allowing individuals to participate in the energy market as producers and consumers. Instead, the EU Regulation 2019/943 \cite{2019/943} sets rules for the electricity market, encouraging participation of consumers and RECs. Additionally, the Regulation on the Governance of the Energy Union and Climate Action (EU 2018/1999 \cite{2018/1999} , updating in 2023) requires member states to develop integrated national energy and climate plans, including measures to support self-consumption groups and RECs.

Italy was the first EU memeber to implement, as laws, the directives of the Clean Energy Package for all European  (\cite{IT1}, \cite{IT3}, \cite{IT4} and  \cite{IT2}).
The Italian energy market is supported by the Italian Energy Services Operator (GSE) and the Italian Regulatory Authority for Energy, Networks and Environment (ARERA). The GSE has been identified by the national government to pursue and achieve environmental sustainability through the two pillars of renewable sources and energy efficiency. ARERA promotes competition and efficiency in public utilities while protecting consumers, balancing economic, social, and environmental objectives to ensure efficient resource use. On January 2024, the Ministry of Environment and Energy Security (MASE) published a decree aimed at enhancing the development of renewable energy, strengthening energy security, and achieving climate goals through the installation of new renewable plants \cite{IT4}. This decree introduces two cumulative measures to promote self-consumption: a nationwide incentive tariff for renewable energy, funded by a surcharge on electricity bills and the ARERA valuation fee, and a non-repayable grant covering up to 40 per cent of eligible costs, financed by the National Recovery and Resilience Plan (NRRP), for communities in municipalities with fewer than 5,000 residents. 
According to the GSE, the incentive tariff, valid for 20 years, is based on installed capacity and market energy prices, while the ARERA valuation fee fluctuates yearly based on the authority's decisions on energy valuation. This regulatory framework will remain in effect until December 2027, or until a total incentivized capacity of 5 GW is achieved, thereby promoting the development of renewable energy and self-sufficiency in smaller communities. Moreover, this incentive tariff covers the energy used to fulfill the power demand of the group and is based on the virtual exchange of energy. In this framework, the self-produced energy is not directly consumed by the group but is sold to the market, while the group's power demand is entirely satisfied by energy purchased from the grid. In this way, by using an electric meter, it is possible to measure how much of the energy took from the grid was actually covered by the self-produced energy sold to the grid, allowing for the application of the incentive tariff.

\begin{figure}[h]
\centering
\includegraphics[scale=0.8]{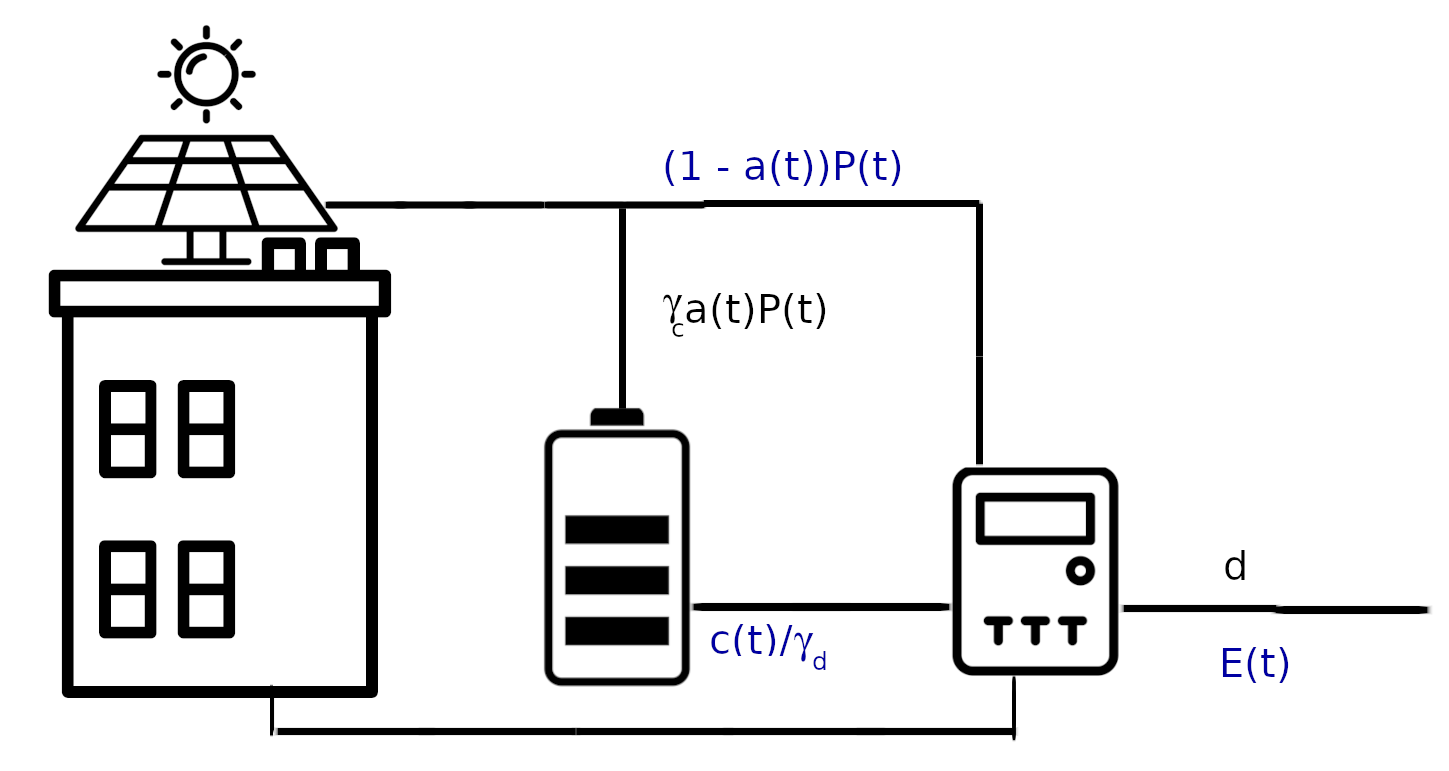} 
\caption{Self-consumption group configuration}
\label{SCG}
\end{figure}

In this paper we concentrate on the optimal management of a self-consumption group which can generate electricity with photovoltaic panels, coupled with a battery storage facility, and sells the surplus electricity on the power market, which receive an incentive tariff according to the Italian framework (virtual exchange of energy). This setting is simpler than a typical REC, as in the latter case various kinds of producers and consumers could be present (see e.g. \cite{AFV} where a biogas producer and photovoltaic prosumers are present). For this reason, while the problem of optimal management of a REC would require a more complex model than this, this model could be a first step in that direction. Our model of self-consumption group assumes that the group operates up to a final horizon $T$ and can produce photovoltaic energy with instantaneous power $P(t)$. The group decides to store a percentage $a(t) \in [0,1]$ of this electricity and sell the remaining power $(1 - a(t)) P(t)$ on the market at the electricity spot price $X(t)$. The stored power can also be sold in the market by discharging the battery at instantaneous power $\gamma(t) \in [0,\Gamma]$. Therefore, the total energy $E(t)$ that is sold to the grid is the sum of the photovoltaic power not stored in the battery plus the power discharged from the battery. A graphical illustration of this scheme is presented in Figure \ref{SCG}. The operations of charging and discharging the battery should be performed in such a way that the total power stored in the battery $S(t)$ remains inside an operational interval $[S_{\mathrm{min}}, S_{\mathrm{max}}]$. 

As stated in the virtual framework, all the power demand $D(t)$ of the group is bought from the grid  and all the energy produced is sold to the grid. The incentive covers the energy used to fulfill the instant power demand. The aim of the self-consumption group is to minimize the total cost up to the final maturity $T$, by optimally charge and discharge the battery.   


Since in this problem, in principle, we have the state variables $X$, $P$ and $D$ which can be random, the natural modeling is that of a stochastic control problem. More in detail, first we assume that $X$, $P$ and $D$ are three multiplicative mean-reverting processes, with a suitable seasonality and possibly with a mutual dependence in the noise, while the state of the battery $S$ has an evolution which depend deterministically on the state $P$ and the charge-discharge control $(\alpha,\gamma)$. 

We choose to tackle this problem by using the dynamic programming principle: this formulation for the problem immediately brings some qualitative results. The first one is that, if the charge and discharge efficiencies are both less than 1 (as it should be physically plausible), then controls where one charge and discharge simultaneously are never optimal. The second one is that the value function is decreasing and convex with respect to the charge state $S$. The dynamic programming principle leads then naturally to a suitable Hamilton-Jacobi-Bellman (HJB) equation, which the value function satisfies in the viscosity sense. This allows us to compute numerically the optimal strategy. In order to get numerical results and help with the interpretation of our findings, we assume that the electricity price $X$ and the power demand $D$ are deterministic functions of time, while the evolution of the photovoltaic production $P$ and of the charge state $S$ remains random. In this simplified setting, we can present closed-form solutions to the problem of finding the optimal charge-discharge strategies. It turns out that, in analogy to other storage problems (see e.g. the seminal paper \cite{TDR} on gas storage), the control is of bang-bang type; more in detail, it is always optimal to restrict the optimization to very few points. In this case, these points are the usual "charge all", "discharge all" and "do nothing", but in addition we have a fourth point, due to the introduction of the self-consumption incentive, which corresponds to charge-discharge so that the demand equals the produced electricity, in order to take full advantage of this incentive. After this theoretical finding, we present numerical results by solving the HJB equation with a mixed explicit-implicit scheme: these results fully confirm our theoretical findings. 

The rest of the paper is organized as follows: Section \ref{review} present the literature review for photovoltaic systems applied to self-consumption groups seen from a legislative, environmental, economical and engineering aspects. Section \ref{general_setting} formalized the problem as a finite horizon stochastic control problem and in Section \ref{general_properties} we present some general properties of the vcontrol problem. Section \ref{two} present the particular case when the electricity price and the power demand are deterministic functions of time. In Section \ref{app} we apply numerical schemes in order to solve the related HJB equations. Section \ref{results} presents the obtained results, where we discuss the optimal battery management at different moments during the day. Section \ref{conclusion} present the conclusion of the paper.

\section{Literature review} \label{review}

Self consumption groups and RECs become a very active and necessary research argument in many different areas, ranging from legislative, environmental and economical aspect, to engineering and physical aspects. From the legislative point of view, researches focuses on the analysis of the normative implementation of EU members to promote creation of this entities, self-producing and consuming their own energy from renewable sources. In \cite{DeLotto} it is presented an overview of the main legislative framework in the EU and Italy in relation to RECs. By analysing the energy consumption of a neighbour in Milan they exemplify the benefit and challenges on implementing this communities, from legislative, social, environmental, economical an technical aspects. Among the benefit are highlighted: the incentive to renewable energy production, reduction in consumption costs, reduction in carbon emissions, life quality improvements and relatively easy management of the community's energy system. In \cite{Canova} it is performed an energy, economic and environmental assessment of creating a self consumption group, considering the incentive scheme applied by the Italian normative. They conclude that self-consumption groups are economically profitable for residential users, with cost savings up to $32\%$ and carbon emission reduction up to $60\%$.

In application to the residential sector, the configuration of the renewable power source of self-consumption groups is mainly based on photovoltaic system, consisting on solar panels and batteries. A review about the optimization methods used for planing photovoltaic systems capacities  in residential applications is presented in \cite{Khezri}. However, the installation costs of this technology not always allows for such configurations without a proper incentive policy, covering part of the costs.  For example, in \cite{Lage} are analyzed the policies on photovoltaic technology in Italy and Portugal for self-consumption groups and RECs. The results show the difference between the policies of both countries, underlying the need of efficient implementation of EU directives. In the Italian framework they found that the best configuration is made by large photovoltaic capacities and smaller batteries. On the other side, in Portugal, only smaller photovoltaic capacities without batteries are profitable. Similar conclusions where obtained in \cite{Shopfer}, where they perform an economical technical assessment of photovoltaic systems in Switzerland. They found that, without any subside or incentive policies, for the $40\%$ of the considered population it was profitable to install photovoltaic panels, but only for the $0.1 \%$  it was profitable to include a battery as storage system. They also found that the profitability of include photovoltaic systems depends on the load demand profile. In \cite{Fang} is analyzed the issue of integrating energy storage systems. In particular they establish that the exciting market mechanisms are not prepare for integration of this technology. They approach this problem by proposing a bidding structure, a corresponding clearing model and a settlement rule. In \cite{BAR} they highlight the need for energy policies to develop market mechanism which facilitate the integration of energy storage. They perform an analysis under collective and individual households scenarios. They found that sharing batteries are $64–94\%$ more effective when the objective is to reduce grid congestion. 

Even though this technology seem to be still too expensive to be applied for residential user without subsides or incentives, under suitable schedules and sizing, batteries contributes with stability to the power systems and power consumption cost reduction. Several authors analyze this issue and propose optimal management strategies in order to obtain the maximum benefit of this systems (see \cite{Danzer} and reference therein). In \cite{Trevisan} and \cite{Zou} are presented studies analyzing economical aspect of optimal management operation and sizing of photovoltaic systems.

Regarding the implementation of photovoltaic systems to self-consumption groups and RECs, in \cite{Mustika} it is studied the management and the profit allocation within an energy community performing collective self-consumption.  They implement the methodology to a case in France, considering deterministic scenarios for the management of a groups consisting in seven households with the combination of photovoltaic panels, batteries, and an electric vehicle. In \cite{GomezG}, it is studied the optimal sizing and power management of household producing and consuming photovoltaic energy. In particular they optimize photovoltaic  power, electric vehicle charging load , household consumption load, battery bank, and power converters. The problem is formulated as an optimization of a deterministic constrain problem in a finite horizon, where the total annual cost of the household microgrid is minimized. This cost includes: initial investment, operation and maintenance, replacement, net income from energy exchange, and net income from frequency containment reserve provision. The formulation is applied to a case in Spain. Among the results, they found that batteries bank are cost-effective way of enhancing the self-consumption of photovoltaic power, by decreasing the consumption costs. When different participants are presented in the production and exchange of energy, as is the case of RECs, among the methodologies it is used game theory since its allows to study equilibrium an found the best strategies under cooperative or competitive frameworks, as it is implemented in \cite{Nagpal}, \cite{MMM}, \cite{Auer}, among others. In \cite{Nagpal} it is applied a hierarchical game where they maximized the self consumption and self production of a local energy community with a share energy storage, composed by three building archetypes, each of them owning photovoltaic panels: residential houses, houses and health care facilities. The optimization is made in two steps: first the end-users minimize their cost by optimally scheduling their loads profile. In a second step, this profile are refined in order to maximize self consumption in a cooperative setting. Future work points out to include stochastic for weather forecast and applied the model to a real case. In \cite{MMM} they apply a cooperative formulation in order to fairly distribute the revenues and cost among the community members. Moreover they perform an optimal allocation of installation of technologies, i. e., the optimal sizing of renewable power generators and energy storage devices.
In \cite{Auer} instead, the configuration of the groups is divided in producer and consumer. The producer owns a distributed energy resources (DER) as photovoltaic power source and batteries, while the consumer contribute with the power demand. They apply a Stackelberg game in order to determine energy allocation and pricing of DER. 
More applications to optimal sizing and management of photovoltaic system can be found \cite{Duman}, \cite{Ntube}, \cite{Tercan}, \cite{Harsha}, \cite{Alhaider} and references therein.

In photovoltaic systems application, ion-lithium batteries are widely used, since their high performance and efficiency. However, lithium extraction is not such a good solution from an environmental point of view. Countries with high reserves of lithium as Australia, Chile, China, and Argentina, suffer the environmental damage of the extraction of this mineral and worldwide, the recycling of the components of this storage devices. See \cite{LiCHL}, \cite{LiARCHL} for detailed analysis of this issue in energy transition. We conclude that optimal sizing and battery management will not also contribute to enhance economical and technical aspect, but also with a more responsible usages of this devices, improving lifetime of the battery and avoiding oversized systems. 

\subsection{Our contribution}

We propose an optimal management strategy for the battery owned by a self consumption group, which produces photovoltaic power from photovoltaic panels. We set the problem as a continuous time finite horizon stochastic control problem, where we consider the variability of photovoltaic production.  We take into account daily seasonal behavior in power demand, electricity price and photovoltaic production in order to obtain a strategy such that the consumption cost is reduced. Our main contribution is to include stochastic processes to model the variability in photovoltaic production. In most research on this topic, the characterization of random components, such as weather or prices, is approached using deterministic functions, the evaluation of specific scenarios or as set of data input.
\section{General setting} \label{general_setting}

{

In this section we formalized the optimal management of a battery as a stochastic optimal control problem. Then, we apply the dynamic programming principle in order to solve the problem with the aid of the associated Hamilton-Jacobi-Bellman equation. 
\\ \indent
To this purpose, we let our system start from time $t \in [0,T]$, and we consider a probability space $(\Omega, \mathcal{F}, \mathbb{P})$ rich enough to support a three-dimensional standard Brownian motion $(W_p,W_x,W_d)$. 

Let us then consider a three-dimensional Ornstein-Uhlenbeck process $(U_x,U_p,U_d)$, satisfying the following stochastic differential equation
\begin{eqnarray}\label{mdou}
\left( \begin{array}{c} dU_x(u)\\
dU_p(u) \\
dU_d(u)
\end{array}  \right) & = & -\left(  \begin{array}{c}
\xi_x  U_x(u) \\
\xi_p U_p(u) \\
\xi_d U_d(u)
\end{array} \right) du + \Sigma \left( \begin{array}{c}
dW_x(u) \\
dW_p(u) \\
dW_d(u)
\end{array}\right), \qquad u \in [t,T],
\end{eqnarray}
\noindent with initial condition $(U_x(t),U_p(t),U_d(t)) = (x,p,d)$.  Here $\xi_x$, $\xi_p$ and $\xi_d$ represent the mean-reversion speeds of $U_x$, $U_p$ and $U_d$, respectively, while the 3 $\times$ 3 matrix $\Sigma$ models the interdependence of the noise components: if we assume that they are independent, it is sufficient to let $\Sigma$ be a diagonal matrix; however, this model can be possibly more general, with interdependencies among the components modeled with non-diagonal terms in $\Sigma$. }

\begin{figure}[h]
     \centering
     \begin{subfigure}[b]{0.45\textwidth}
         \centering
         \includegraphics[width=\textwidth]{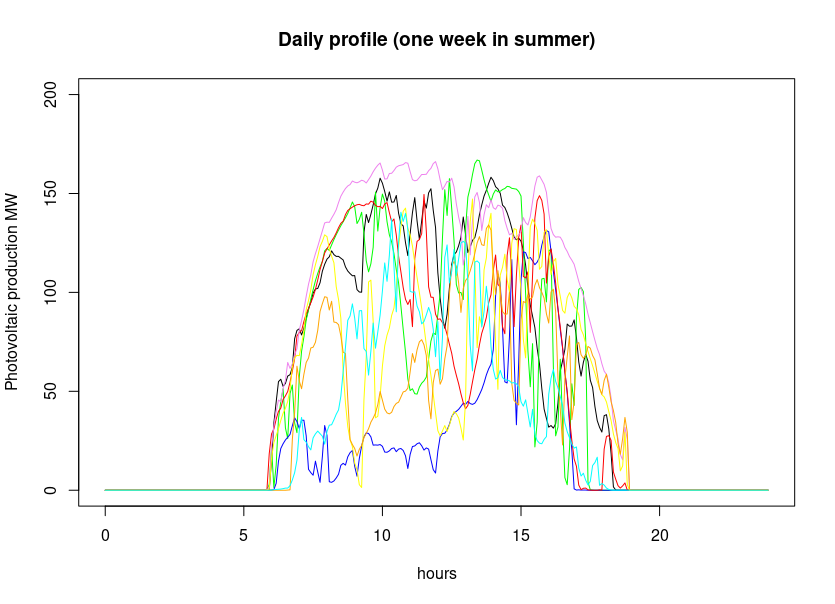}
         \caption{Photovoltaic production during one week in summer}
         \label{phs}
     \end{subfigure}
     \begin{subfigure}[b]{0.45\textwidth}
         \centering
         \includegraphics[width=\textwidth]{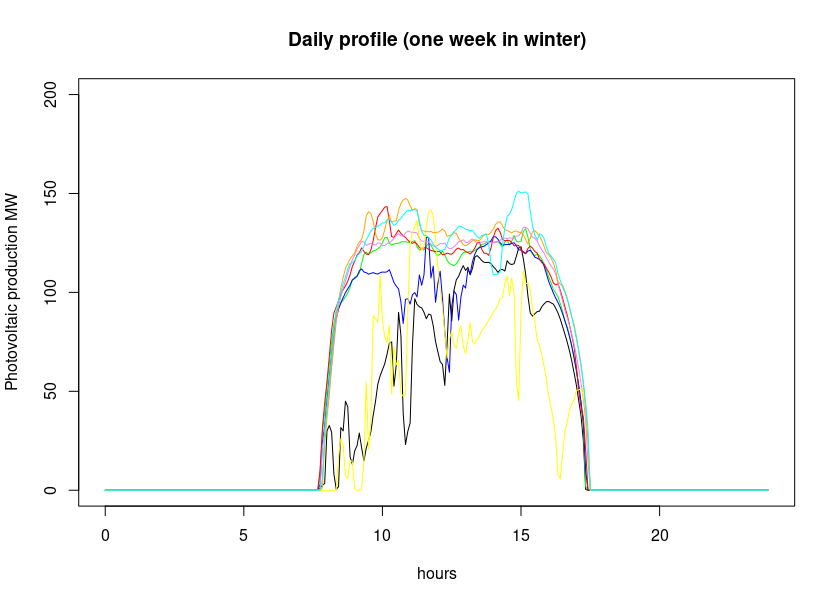}
         \caption{Photovoltaic production during one week in winter}
         \label{phw}
     \end{subfigure}
        \caption{Photovoltaic production profiles. The plotted data was obtained from \cite{Data_photo}, and correspond to forecast measures every 5 minutes in a location in Florida (USA).}
        \label{photo}
\end{figure}

We consider a self-consumption group, which owns a photovoltaic capacity $Y$ and receives a unitary incentive $Z$ for the shared energy. We will also suppose that they own a storage device which they aim to  manage optimally in order to minimize their power consumption costs. We denote by $P(t)$ the instantaneous  photovoltaic power production of the group at time $t$, which we define as 
\begin{eqnarray}
   P(u) = f_p(u) e^{U_p(u)}, \qquad u \in [t,T], 
   \label{Pp}
\end{eqnarray}
\noindent with expected value
\begin{eqnarray} \label{EPp}
      \mathbb{E}\left[  P(u)\right] =  A \sin( 2 \pi \psi (u + \phi))e^{p e^{\xi_p(t - u) }} e^{\frac{\sigma_p^2}{4\xi_p} \left(1 - e^{2\xi_p(t - u)} \right) }, 
\end{eqnarray}
where the function $f_p$ represents the daily seasonal behavior of power production. {To give an idea of how $f_p$ can be modeled,} in Figure \ref{photo} we show examples of different photovoltaic production profiles. 
{Though in Figure \ref{photo} yearly seasonal factors seem present and can possibly be incorporated in industrial applications, for the sake of simplicity here we choose to model only daily seasonality; thus, w}e will represent the daily seasonality of the photovoltaic production by setting
\begin{eqnarray}
f_p(u) = A \max\{ \sin(\pi \psi (u +\phi)), 0\} ,
\label{Ps}
\end{eqnarray}
\noindent where $\psi = 1/24$h,  $\phi$ is the daily phase and  $A$ (MW) the amplitude. 

The produced power $P(t)$ can be either sold immediately in the electricity market at the electricity spot price, denoted by $X(t)$, or stored in a device. If sold, we assume that the electricity spot price evolves according to
\begin{eqnarray}
    X(u) = f_x(u) { e^{U_x(u)}}, \qquad u \in [t,T]
    \label{Xs}
\end{eqnarray}
\noindent  where $f_x > 0$ is a deterministic function describing the daily seasonality of the electricity spot price. Again, for simplicity we are not considering monthly or yearly seasonalities. Moreover,  notice that we are implicitly assuming that the electricity price is strictly positive. 

\begin{figure}[h]
     \centering
     \begin{subfigure}[b]{0.45\textwidth}
    \centering
    \includegraphics[scale = 0.38]{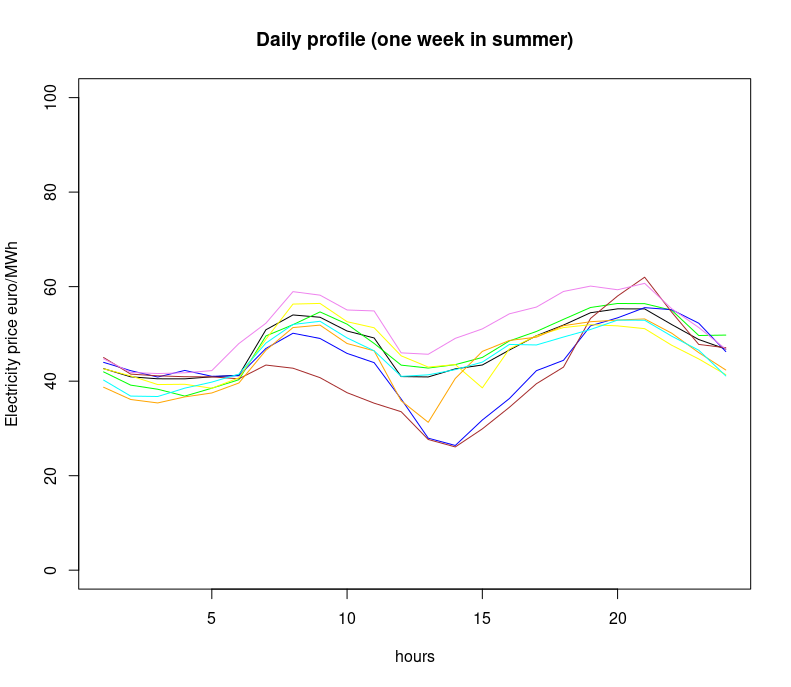}
    \caption{Daily profile of the electricity price in summer}
    \label{ED_season_s}
     \end{subfigure}
     \begin{subfigure}[b]{0.45\textwidth}
    \centering
    \includegraphics[scale = 0.38]{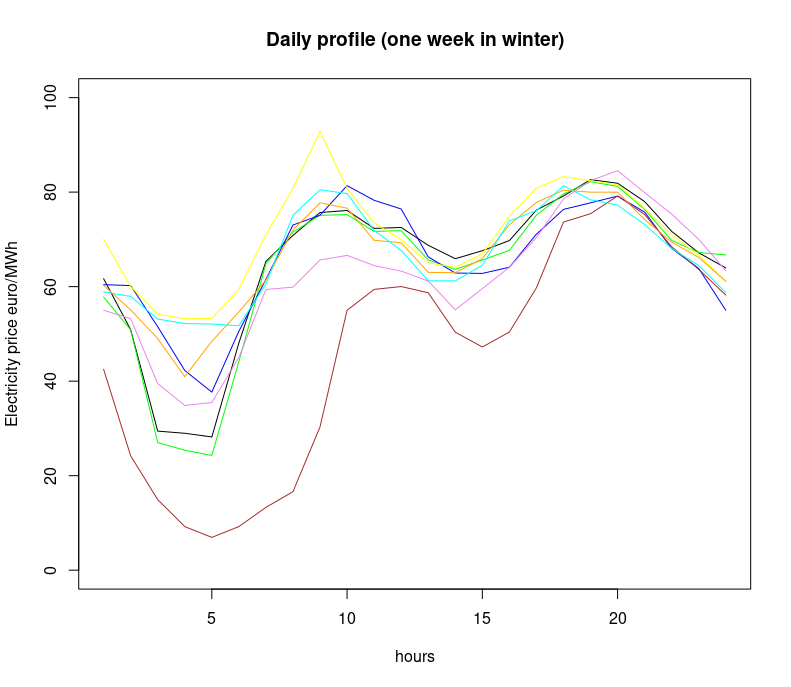}
    \caption{Daily profile of the electricity price in winter}
    \label{ED_season_w}
     \end{subfigure}
        \caption{Daily electricity price profile.}
        \label{ED_season}
\end{figure}


We denote by $S(u)$ the state of charge of the battery at time $u$, which is assumed to be controlled through a charge action $\alpha$, representing the proportion of the photovoltaic power that we want to store, and a discharge action $\gamma$, representing the quantity of the stored power that we want to sell. We denote by $\alpha_{max}$ the maximum charging power of the battery and by $\Gamma$ the maximum discharging power of the battery. 
 
Assuming that $S(t) = s$, we let $S$ evolve as
{
\begin{eqnarray} \label{S}
S(u) = s +  \eta_c \int_t^u \alpha(v) P(v) dv - \frac{1}{\eta_d} \int_t^u \gamma(v) dv, \qquad u \in [t,T], 
\end{eqnarray}
\noindent where $\eta_c \leq 1$ is the charge efficiency and $\eta_d \leq 1$ is the discharge efficiency of the storage device. The state of charge $S$ lies inside the operational values of the battery $S_{\mathrm{max}}$ and  $S_{\mathrm{min}}$. Similar models for batteries can be found in \cite{BAR} or \cite{MMM}.
\footnote{for simplicity we are not considering the wear suffered by the battery. 
Actually, the more the changes in charge-discharge strategies, the higher is the wear in the battery. We assume the constraint $S_{min}$ and $S_{max}$ such that the wear suffered by the battery is negligible.
} 
\cite{11}. Therefore we have the state constraint for the state of charge:
\begin{eqnarray*}
S_{\mathrm{min}} \leq S(u) \leq S_{\mathrm{max}}, \hspace{1cm} u \in [t,T].
\label{constr}
\end{eqnarray*}
Let us call the pair $(\alpha, \gamma)$ the storing management strategy, and call $\mathcal{A}(t,s)$ the set of {admissible charge and discharge strategies for a storage level starting from $S(t) = s$, defined by}

$$\mathcal{A}(t,s) := \left\{ (\alpha, \gamma): \Omega \times [t, T] \xrightarrow{} [0,1] \times [0,\Gamma] \left| \begin{array}{c} 
\alpha, \gamma \mbox{ are progressively measurable processes such that }\\
\alpha P(t) \leq \alpha_{max} \mbox{ and }
 S(u) \in [S_{\mathrm{min}},S_{\mathrm{max}}]\ \forall u \in [t,T] \\
\end{array}
\right. \right\}.$$

\noindent 
}

Finally, we suppose that the power demand of the self-consumption group is
\begin{eqnarray}
    D(u) = f_d(u) e^{U_d(u)}, \qquad u \in [t,T], 
    \label{Ds}
\end{eqnarray}
\noindent where $U_d$ is an Ornstein-Uhlenbeck process and $f_p$ a strictly positive function, describing the daily seasonal consumption profile of a residential demand, just as in the previous description of the state variable we are not considering monthly or yearly seasonalities. Figure \ref{power_demand} shows daily profiles in summer and winter of power demand in Italy.


\begin{figure}[h]
     \centering
     \begin{subfigure}[b]{0.45\textwidth}
     \centering
    \includegraphics[scale =0.38] {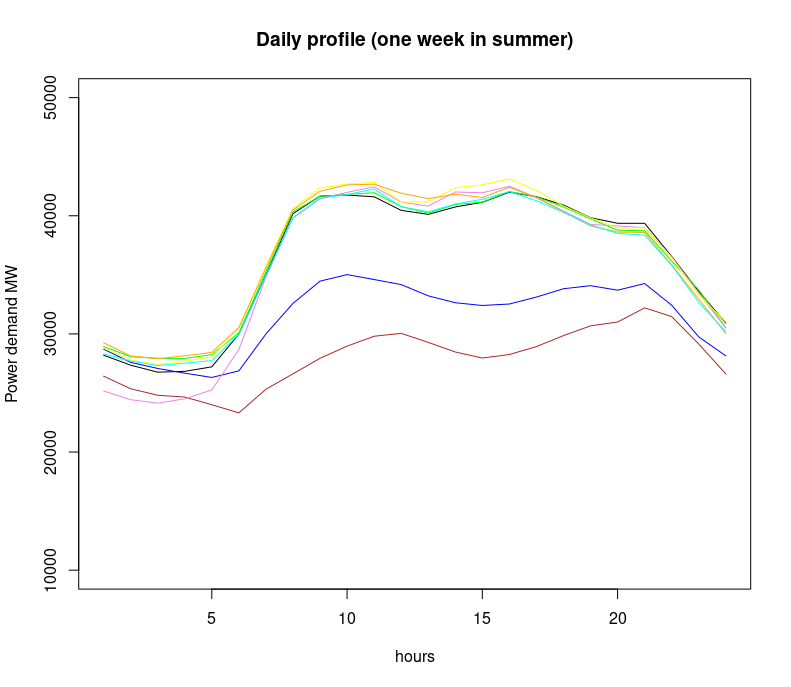}
    \caption{Power demand daily profile in summer}
    \label{D_season}
     \end{subfigure}
     \begin{subfigure}[b]{0.45\textwidth}
     \centering
    \includegraphics[scale =0.38] {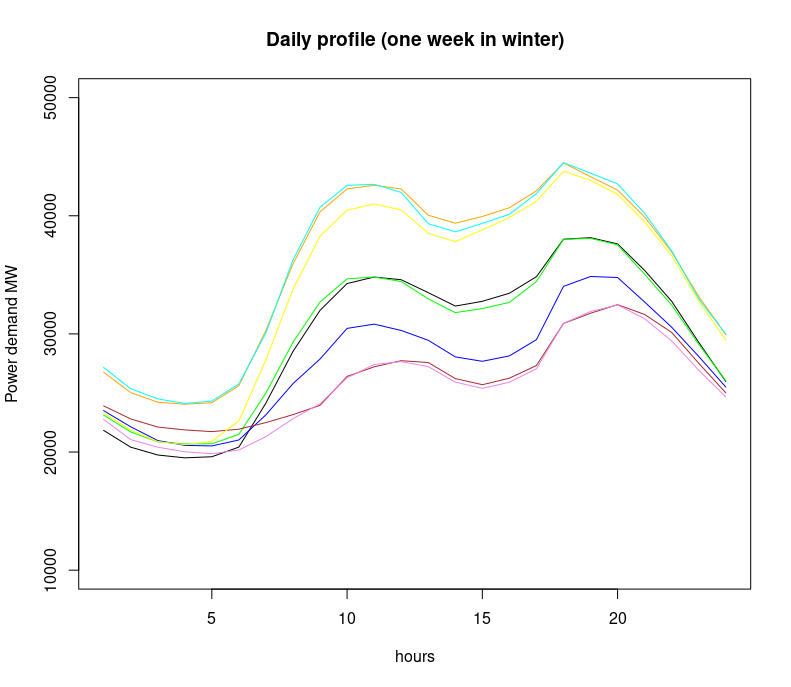}
    \caption{Power demand daily profile in winter}
    \label{D_season_w}
     \end{subfigure}
        \caption{Daily power demand profile.}
        \label{D_season_global}
\end{figure}

According to the European normative, the union members have to implement in their legislation some mechanisms in order to incentives renewable energy production. One of the first application as a law in the country members is the Italian case. In the Italian normative, there is a unitary incentive $Z$ in \euro/MWh for the self-consumed power of the group which is defined, at time $u$, as the minimum between the instantaneous power demand and the instantaneous self-produced energy, i.e. 
\begin{eqnarray*}
    \min \{ D(u) , E(u) \},
\end{eqnarray*}
\noindent where $E(u)$ is the instantaneous power sold to the grid at time $u \in [t,T]$, which is the part of the photovoltaic production which is not stored, plus the effective power discharged from the battery, i.e.,
\begin{eqnarray}
    E(u) = (1 - \alpha(u) ) P(u) +  \gamma(u) 
\end{eqnarray}
The power consumption cost associated to the group administrator is given by  the total cost of the electricity bought from the power grid, net of self-production, minus the total incentive, with both of these quantities being discounted with a free risk discount rate $r > 0$ and computed in the time interval $[t,T]$. Thus, the cost associated to a charge-discharge strategy $(\alpha,\gamma) \in {\mathcal A}(t,s)$ is

\begin{equation}\label{J}
J(t, x,d,p,s, \alpha, \gamma ) := \mathbb{E} \left[ \int_{t}^{T} e^{- r u} X(u) ( D(u) - E(u))du \right] - Z\mathbb{E} \left[ \int_t^{T} e^{-ru} \min\{ D(u), E(u) \} du \right].
\end{equation}
In order to make more explicit the dependence of $J$ on the state variables $x$, $d$ and $p$, let 

$$ h_1(u,x,d,p,a,c) = e^{- r u}(f_x(u) e^x)\left(f_d(u) e^{d} - \left((1 - a ) f_p(u) e^{p} + c \right) \right), $$

represent the difference between the energy that the group buys in order to satisfy its demand and the energy that the group decides to sell in the electricity market, and 

$$ h_2(u,x,d, p, a, c) = e^{-ru} \min\left\{ f_d(u) e^{d}, (1 - a) f_p(u) e^{p} + c \right\} $$

the incentive received by the community. So, the total discounted running cost $h$  is
\begin{equation} \label{h}
h(u,x,d, p, a, c) := h_1(u,x,d, p, a, c) - Z h_2(u,x,d, p, a, c) 
\end{equation}
and we can rewrite the cost $J$ as
\begin{equation} \label{J2}
J(t, x,d,p,s, \alpha, \gamma ) := \mathbb{E} \left[ \int_{t}^{T} h(u,U_x(u),U_d(u),U_p(u),\alpha(u),\gamma(u))  du \right].
\end{equation}


We assume the presence a group administrator, which aims to manage optimally the storage device in order to minimize the consumption costs of the group, i.e., find a strategy $(\alpha^*, \gamma^*) \in \mathcal{A}(t,s)$ such that

\begin{eqnarray}
J(t,x,d,p,s,\alpha^*, \gamma^*) = \min_{(\alpha, \gamma) \in \mathcal{A}(t,s)} J(t,x,d,p,s,\alpha, \gamma).
\end{eqnarray}

\noindent We define the value function of our optimal control problem as 

\begin{eqnarray} \label{value4}
V(t,x,d,p,s) :=  \inf_{(\alpha, \gamma) \in \mathcal{A}(t,s)} J(t,x,d,p,s,\alpha, \gamma). 
\label{V}
\end{eqnarray}

\section{General properties of the control problem}
\label{general_properties}

{
The first that we show is that controls which have both  components $(\alpha,\gamma)$ nonzero are never optimal. The idea is that, if $\alpha(u) \gamma(u) > 0$, then we can decrease both $\alpha$ and $\gamma$ in Equation \eqref{S} so that the process $S$ is unchanged. 

\begin{proposition} \label{zero}
If $(\alpha,\gamma) \in {\cal A}(t,s)$ is such that ${\mathbb E}[\int_t^T \alpha(u) \gamma(u) du] > 0$, then there exists $(\alpha',\gamma') \in {\cal A}(t,s)$ such that ${\mathbb E}[\int_t^T \alpha'(u) \gamma'(u) du] = 0$ and $J(t,x,d,p,s,\alpha',\gamma') \leq J(t,x,d,p,s,\alpha,\gamma)$. This inequality is strict if $\eta_c \eta_d < 1$ and $f_x(u) > 0$ for all $u \in [t,T]$.
\end{proposition}
\begin{proof}
For all $u \in [t,T]$ define 
$$ (\alpha',\gamma')(u) = \left\{ \begin{array}{ll}
(\alpha,\gamma)(u)  & \mbox{ if } \alpha (u) \gamma(u) = 0, \\
\\
(0,\gamma(u) - \eta_c \eta_d \alpha(u) P(u))    & \mbox{ if } \alpha (u) \gamma(u) > 0 \mbox{ and } \eta_c \alpha(u) P(u) - \frac{\gamma(u)}{\eta_d} \leq 0,\\
\\
(\alpha(u) - \frac{\gamma(u)}{\eta_c \eta_d P(u)},0)    & \mbox{ if } \alpha (u) \gamma(u) > 0 \mbox{ and } \eta_c \alpha(u) P(u) - \frac{\gamma(u)}{\eta_d} > 0.\\
\end{array} \right. $$
Then ${\mathbb E}[\int_t^T \alpha'(u) \gamma'(u) du] = 0$; moreover, if we call $E'(u) := (1 - \alpha'(u)) P(u) + \gamma'(u)$, we have
\begin{eqnarray*}
E'(u) - E(u) & = & (\alpha(u) - \alpha'(u)) P(u) + \gamma'(u) - \gamma(u) = \\
& = & \left\{ \begin{array}{ll}
0  & \mbox{ if } \alpha (u) \gamma(u) = 0, \\
\\
\alpha(u) P(u) (1 - \eta_c \eta_d)    & \mbox{ if } \alpha (u) \gamma(u) > 0 \mbox{ and } \eta_c \alpha(u) P(u) - \frac{\gamma(u)}{\eta_d} \leq 0,\\
\\
\gamma(u) (1 - \frac{1}{\eta_c \eta_d})   & \mbox{ if } \alpha (u) \gamma(u) > 0 \mbox{ and } \eta_c \alpha(u) P(u) - \frac{\gamma(u)}{\eta_d} > 0.\\
\end{array} \right.
\end{eqnarray*}
Thus, $E'(u) \geq E(u)$, and also $\min(D(t),E'(t)) \geq \min(D(t),E(t))$. It follows that
\begin{eqnarray*}
\lefteqn{J(t,x,d,p,s,\alpha',\gamma') - J(t,x,d,p,s,\alpha,\gamma) = } \\
& = & - \mathbb{E}[\int_t^T e^{-ru} X(u) (E'(u) - E(u)) du] - Z \mathbb{E}[\int_t^T e^{-ru} \min(D(t),E'(t)) - \min(D(t),E(t))) du] \leq \\
& \leq & - \mathbb{E}[\int_t^T e^{-ru} X(u) (E'(u) - E(u)) du] \leq 0
\end{eqnarray*}
and the inequality is strict if $\eta_c \eta_d < 1$ and $f_x(u) > 0$ for all $u \in [t,T]$.
\end{proof}
}
\begin{lemma} \label{lemma_c}
The value function $V$, defined in Equation \eqref{value4}, is decreasing and convex in $s \in [0,\bar S]$ for all fixed $t$, $x$, $d$, $p$.
\end{lemma}
\begin{proof}
In order to prove monotonicity, take $s_1 < s_2$ such that $s_1, s_2 \in [0,\bar S]$. Then, for all $(\alpha,\gamma) \in {\cal A}(t,s_1)$ we can define
$$ \tau := \inf\left\{ s \in (t,T]\ |\ s_2 - \Gamma (s - t) < s_1 + \eta_c  \int_t^s \alpha(u) P(u) du - \frac{1}{\eta_d} \int_t^s \gamma(u) du \right\} $$
and the modified strategy $(\alpha',\gamma') \in {\cal A}(t,s_2)$ as
$$ (\alpha'(u),\gamma'(u)) := \begin{cases}
(0,\Gamma), & u \in [t,\tau), \\
(\alpha(u),\gamma(u)),  & u \in [\tau,T]. 
\end{cases} $$
so that we have $S^{s_1,\alpha,\gamma}(u) < S^{s_2,\alpha',\gamma'}(u)$ for all $u < \tau$ and $S^{s_1,\alpha,\gamma}(u) = S^{s_2,\alpha',\gamma'}(u)$ for all $u \geq \tau$. This implies 
\begin{eqnarray*}
\lefteqn{ J(t,x,s_1,d,p,\alpha,\gamma) - J(t,x,s_2,d,p,\alpha',\gamma') = } \\
& = & \mathbb{E} \left[ \int_{t}^{\tau} e^{-ru} [\alpha(u) X(u) P(u) + X(u) (\Gamma - \gamma(u)) + \right. \\
&   & + \left. Z (\min(D(u),P(u) + \frac{\Gamma}{\eta_d}) - \min(D(u),(1 - \alpha(u))P(u) + \frac{\gamma(u)}{\eta_d})] du \right] \geq 0 
\end{eqnarray*}
meaning $J(t,x,s_1,d,p,\alpha,\gamma) \geq J(t,x,s_2,d,p,\alpha',\gamma') \geq V(t,x,s_2,d,p)$ as $(\alpha',\gamma') \in {\cal A}(t,s_2)$. Since this holds for all $(\alpha,\gamma) \in {\cal A}(t,s_1)$, we obtain $V(t,x,s_1,d,p) \geq V(t,x,s_2,d,p)$. 

Now we prove convexity. For this, consider $s_1 < s_2$ and, for all $\epsilon > 0$, take two $\epsilon$-optimal strategies $(\alpha_i,\gamma_i)$, $i = 1,2$, i.e. $(\alpha_i,\gamma_i) \in {\cal A}(t,s_i)$ such that $J(t,s_i,p,\alpha_i,\gamma_i) < V(t,s_i,p) + \epsilon$. Define now, for all $\lambda \in [0,1]$, the convex combinations $s_\lambda := \lambda s_1 + (1 - \lambda) s_2$ and $(\alpha_\lambda,\gamma_\lambda) := \lambda (\alpha_1,\gamma_1) + (1 - \lambda) (\alpha_2,\gamma_2)$; then, since the dynamics of the state $S$ is linear in the controls, we have $(\alpha_\lambda,\gamma_\lambda) \in {\cal A}(t,s_\lambda)$. Moreover, since the cost function $h$ is convex in $(\alpha,\gamma)$, we have 
\begin{eqnarray*}
\lefteqn{ V(t,s_\lambda,p) \leq J(t,s_\lambda,p,\alpha_\lambda,\gamma_\lambda) = \mathbb{E} \left[ \int_{t}^{T} h(u,U_x(u),U_d(u),U_p(u),\alpha_\lambda(u),\gamma_\lambda(u))  du \right] \leq } \\
& \leq & \mathbb{E} \left[ \int_{t}^{T} [\lambda h(u,U_x(u),U_d(u),U_p(u),\alpha_1(u),\gamma_1(u)) + (1 - \lambda) h(u,U_x(u),U_d(u),U_p(u),\alpha_2(u),\gamma_2(u))] du \right] = \\
& = & \lambda J(t,s_1,p,\alpha_1,\gamma_1) + (1 - \lambda) J(t,s_2,p,\alpha_2,\gamma_2) < \lambda V(t,s_1,p) + (1 - \lambda) V(t,s_2,p) + 2 \epsilon
\end{eqnarray*}
Since this holds for all $\epsilon > 0$, we have $V(t,s_\lambda,p) \leq \lambda V(t,s_1,p) + (1 - \lambda) V(t,s_2,p)$, i.e. $V$ is convex in $s$. 
\end{proof}

\begin{lemma} \label{lemma_c}
The value function $V$, defined in Equation \eqref{value4}, is decreasing  in $p \in \R$ for all fixed $t$, $x$, $s$, $d$.
\end{lemma}
\begin{proof}
In order to prove the statement, we reparameterize the space of controls, by taking the equivalent formulation 
$$ S'(u) = s +  \eta_c \int_t^u \alpha'(v) dv - \frac{1}{\eta_d} \int_t^u \gamma'(v) dv, \qquad u \in [t,T], $$
where again we assume that $S'(u) \in [S_{\mathrm{min}}, S_{\mathrm{max}}]$, and the control $(\alpha',\gamma')$ is now assumed to belong to the set of admissible controls
$$\mathcal{A}'(t,p,s) := \left\{ (\alpha', \gamma'): \Omega \times [t, T] \xrightarrow{} [0,\alpha_{\mathrm{max}}] \times [0,\Gamma] \left| \begin{array}{c} 
\alpha', \gamma' \mbox{ are progressively measurable processes such that }\\
\alpha'(t) \leq P_p(t) \mbox{ and }
 S'(u) \in [S_{\mathrm{min}},S_{\mathrm{max}}]\ \forall u \in [t,T] \\
\end{array}
\right. \right\}.$$
where we explicitly indicate the dependence on $p$ both of the space of controls $\mathcal{A}'$ and of $P_{p}(u) := f_p(u) e^{U_{p}(u)}$, $u \in [t,T]$. 
In this way, $\alpha'$ is now the photovoltaic energy intensity that we store, and not its proportion as $\alpha$ is. All the other state variables are unchanged, except for the instantaneous power sold to the grid, which now is
$$ E'(u) := P(u) - \alpha(u) + \gamma(u), \qquad u \in [t,T]. $$
The cost $J'$ associated to the strategy $(\alpha',\gamma')$ follows Equation \eqref{J}, by substituting $E$ with $E'$; there, it is immediate to see that $J'$ is decreasing with respect to the process $E'$, in the sense that if $E'_1(u) \leq E'_2(u)$ for all $u \in [t,T]$, then the reverse inequality follows for $J'$. Finally, we notice that these two formulations are equivalent, in the sense that 
$$ V(t,x,s,d,p) = \inf_{(\alpha',\gamma') \in \mathcal{A}'(t,p,s)} J'(t,x,s,d,p,\alpha',\gamma') = \inf_{(\alpha,\gamma) \in \mathcal{A}(t,s)} J(t,x,s,d,p,\alpha,\gamma) $$
We now proceed with the proof. First notice that, if $p_1 \leq p_2$, then elementary properties of Ornstein-Uhlenbeck processes imply that $U_{p_1}(u) \leq U_{p_2}(u)$ for all $u \in [t,T]$, and consequently also 
$$ P_{p_1}(u) := f_p(u) e^{U_{p_1}(u)} \leq P_{p_2}(u) := f_p(u) e^{U_{p_2}(u)} \qquad \mbox{ for all } u \in [t,T] $$
Since the process $P$ is nonnegative, this implies that, if $p_1 \leq p_2$, then $\mathcal{A}'(t,p_1,s) \subseteq \mathcal{A}'(t,p_2,s)$. 
Take now $p_1 \leq p_2$, and consider any given  $(\alpha',\gamma') \in \mathcal{A}'(t,p_1,s)$; then we will also have that 
$(\alpha',\gamma') \in \mathcal{A}'(t,p_2,s)$, and 
$$ J'(t,x,s,d,p_1,\alpha',\gamma') \geq J'(t,x,s,d,p_2,\alpha',\gamma') $$
This implies
\begin{eqnarray*} 
V(t,x,s,d,p_1) & = & \inf_{\mathcal{A}'(t,p_1,s)} J'(t,x,s,d,p_1,\alpha',\gamma') \geq \inf_{\mathcal{A}'(t,p_1,s)} J'(t,x,s,d,p_2,\alpha',\gamma') \geq \\
& \geq & \inf_{\mathcal{A}'(t,p_2,s)} J'(t,x,s,d,p_2,\alpha',\gamma') = V(t,x,s,d,p_2) 
\end{eqnarray*}
where the last inequality follows from $\mathcal{A}'(t,p_1,s) \subseteq \mathcal{A}'(t,p_2,s)$. This concludes the proof. 
\end{proof}

Let us move into the characterization of the value function \eqref{V}. We use the dynamic programming approach in order to describe the value function \eqref{V} as a solution of the Hamilton Jacobi Bellman equation. By applying dynamic programming principle and standard calculations we obtain the following elliptic partial differential equation

\begin{eqnarray}\label{HJB}
& &-\frac{\partial V(t,x,d,p,s)}{\partial t}  - \inf_{(a,c) \in \mathcal{A}_p(s)} \left[ \mathcal{L}^{a,c} V(t,x,d,p,s) + h_1(t,x,d,p,a,c) -Z  h_2(t,d,p,a,c) \right] =0,
\end{eqnarray}
\noindent where
\begin{eqnarray}\label{gL}
\mathcal{L}_t^{a,c} V(t,x,d,p,s)& =& b(t,x,d,p,s,a,c) D_{x,d,p,s}V(t,x,d,p,s) + \frac{1}{2} \tr \left( \sigma \sigma'D_{x,d,p,s}^2 V(t,x,d,p,s) \right),
\end{eqnarray}
\noindent where

$$ b(t,x,d,p,s,a,c) = \left( -\xi_x x, -\xi_p p , -\xi_d d, \eta_c a f_p(t)e^p - \frac{c}{\eta_d}  \right) , \qquad  \sigma = 
\left( \begin{array}{c c} 
\Sigma & 0\\
0 & 0
\end{array} \right) $$

and 
\begin{equation} \label{Ap}
\mathcal{A}_p(s) := \left\{ \begin{array}{ll}
    [0,1] \times \{0\}      & \mbox{ if } s = S_{min},\\
    {[0,1] \times [0,\Gamma]} & \mbox{ if } s \in (S_{min},S_{max}),\\
    \{0\} \times [0,\Gamma] & \mbox{ if } s = S_{max}. \\
    \end{array} \right. 
\end{equation}
In the case that $\Sigma = \mathrm{diag}(\sigma_x,\sigma_p,\sigma_d)$, \eqref{gL} is written as
\begin{eqnarray*}
\mathcal{L}_t^{a,c} V(t,x,d,p,s) & = & -\xi_x x V_x + -\xi_d d V_d + -\xi_p p  V_p + \left(\eta_c a f_p(t)e^p - \frac{c}{\eta_d} \right) V_s\\
& & + \frac{1}{2} \left( \sigma_x^2 V_{xx} + \sigma_d^2 V_{dd} + \sigma_p^2 V_{pp} \right),
\end{eqnarray*}

Finally, equation \eqref{HJB} is written as
\begin{gather} \label{HJB}
\begin{cases}
\displaystyle  -\frac{\partial V}{\partial t} +\xi_x x V_x + \xi_d d V_d + \xi_p p  V_p  - \frac{1}{2} \left( \sigma_x^2 V_{xx} + \sigma_d^2 V_{dd} + \sigma_p^2 V_{pp} \right) = \\
 \inf_{(a,c) \in \mathcal{A}_p} \left[\left(\eta_c a f_p(t)e^p - \frac{c}{\eta_d}\right)V_s  + h_1(t,x,d,p,a,c) -Z  h_2(t,d,p,a,c) \right]\\
V(T,x,p,d,s) = 0, \hspace{0.5cm} \forall (x,p,d,s) \in \mathbb{R} \times [0, y_h]\times \mathbb{R}^+ \times [S_{min}, S_{max}].
\end{cases}
\end{gather}

The connection between the value function $V$ defined in Equation \eqref{value4} and the Hamilton-Jacobi-Bellman equation \eqref{HJB} is classically formulated via a verification theorem. However, in our case the second derivative in $s$ is missing, thus we cannot expect to have classical solutions, i.e. with continuous second derivatives. An additional complication is that our equation is state-constrained: in fact, the state variable $S$ is not allowed to leave the operational interval $[S_{min},S_{max}]$, and the set of controls $\mathcal{A}_p$ has been defined exactly for that purpose. For this reason, we will present the connection between $V$ and Equation \eqref{HJB} by using the concept of viscosity solutions.

\begin{theorem} \label{visc}
The value function $V$ defined in Equation \eqref{value4} is the unique continuous viscosity solution of the state-constrained Hamilton-Jacobi-Bellman equation \eqref{HJB}.
\end{theorem}

The proof follows by the dynamic programming and comparison principle for viscosity solutions of state-constrained HJB equations. We refer the interested reader to \cite{BarBou} and \cite{CapDolLio} for further details.

{
\section{Photovoltaic production and state of charge as state variables} \label{two}

In this section we study the particular case when the electricity price $X(u)$ and the power demand $D(u)$ are time dependent deterministic functions, while the photovoltaic production $P(u)$ and the state of charge $S(u)$ evolves randomly as in the previous section. This corresponds to letting
$$ \xi_x = \xi_d = 0, \qquad \mbox{ and } \qquad \Sigma := \left( \begin{array}{c c c}
0 & 0 & 0\\
0 & 0 & 0 \\
0 & 0 & \sigma_p 
\end{array} \right) , $$
in \eqref{mdou}, so that $U_x(u) \equiv x$ and $U_d(u) \equiv d$ for all $u \in [t,T]$.

Let us call 
$$\hat{h}_1(t,p,a,c) =  e^{-rt}(f_x(t) e^x)\left(f_d(t)e^d  - (1 - a)f_p(t) e^{p} - c \right) $$ 
\noindent and 
$$\hat{h}_2(t,p,a,c) =  e^{-rt}\min\left\{f_d(t)e^d, (1 - a)f_p(t) e^{p} + c\right\}. $$ 

\noindent Then, the Hamilton-Jacobi-Bellman equation is written as
\begin{gather} \label{HJB2}
\begin{cases}
\displaystyle -\frac{\partial V}{\partial t}  +\xi_p p  V_p  - \frac{\sigma_p^2}{2}   V_{pp}  = 
 \inf_{(a,c) \in \mathcal{A}_p} \left[ \left(\eta_c a f_p(t)e^p - \frac{c}{\eta_d} \right)V_s  + \hat{h}_1(t,p,a,c) -Z  \hat{h}_2(t,p,a,c) \right],\\
 V(T,p,s) = 0, \hspace{1cm} \forall (p,s) \in [0, y_h] \times [S_{min}, S_{max}].
\end{cases}
\end{gather}

The following result is a particular case of Theorem \ref{visc}.

\begin{theorem}
The value function $V$ defined in Equation \eqref{value4} is the unique continuous viscosity solution of the state-constrained Hamilton-Jacobi-Bellman equation \eqref{HJB2}.
\end{theorem}
 
\subsection{Internal solutions} \label{small}

Since it is not possible to obtain a explicit solution for the HJB \eqref{HJB2}, in this section we compute the optimal management for the battery for states $s \in (S_{min}, S_{max})$, i.e., we exclude the case when the battery is fully charge or fully discharge. As the charge and discharge actions are performed with finite velocity we can assume that any strategy, $\alpha \in [0, 1]$, $\gamma \in [0, \Gamma]$, is admissible for values of the state of charge inside the interval $s \in (S_{min}, S_{max})$, since any of these actions will maintain the state of charge $s$ inside the operation interval of the battery. Hence, for this particular case we obtain explicit expression for the optimal strategy.

\begin{figure}[h]
    \centering
   \includegraphics[scale=0.4]{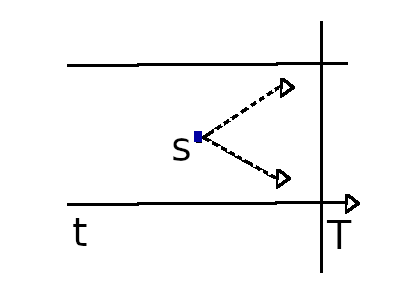}
    \caption{Small time to maturity}
    \label{sttm}
\end{figure}
{
We now formulate the minimization problem in Equation \ref{HJB2} using the shorthand notation
$$ P := f_p(t) e^p, \qquad X := f_x(t) e^x, \qquad D := f_d(t) e^d, \qquad E := (1 - a) P + c, $$
In this way, the quantities $P$, $X$ and $D$ are independent of the control $(a,c)$ but rather depend on state variables $t$, $p$, $x$ and $d$ (these two latter being held fixed to given constants in this section), while $E$ depends linearly on $(a,c)$. Thus, we have to solve}
\begin{eqnarray}
    \inf_{(a,c) \in \mathcal{A}_p} \left[ \left(\eta_c a P - \frac{c}{\eta_d} \right)V_s  + e^{-rt} X (D - E ) -e^{-rt} Z \min\{D, E \} \right]
    \label{inf}
\end{eqnarray}
Observe that the infimum is taken over a piecewise linear function, in the sense that when $D \leq E$ 
we will have a certain linear function in $a$ and $c$, and {we will have another linear expression when $D > E$}
.Therefore, we have to distinguish between these two cases and consider the feasible strategy in every region. 
{
In Figure \ref{figDE} we present how the control region $\mathcal{A}_p = [0,1] \times [0,\Gamma]$ is divided in these two regions, separated by $D = E$, i.e. 
\begin{gather}
c = a P + D - P , 
\label{line}
\end{gather}
More in detail, in Figure \ref{figDE_1} we present the case when $P = 0$, while in Figure \ref{figDE_2} we present one of the possible cases when $P > 0$: in both cases, the white region correspond to case $D \leq E$, while the grey region corresponds to the case $D \geq E$. Finally, thanks to Proposition \ref{zero}, we can restrict the search of minima in a part of the boundary of $\mathcal{A}_p$, namely the left and bottom side in $\partial^- \mathcal{A}_p := ([0,1] \times \{0\}) \cup (\{0\} \times [0,\Gamma]$). 
}

%

\begin{figure}[H]
    \centering
    \begin{subfigure}[b]{0.4\textwidth}
    \centering
\href{cases.webm}{
\includegraphics[scale=0.35]
{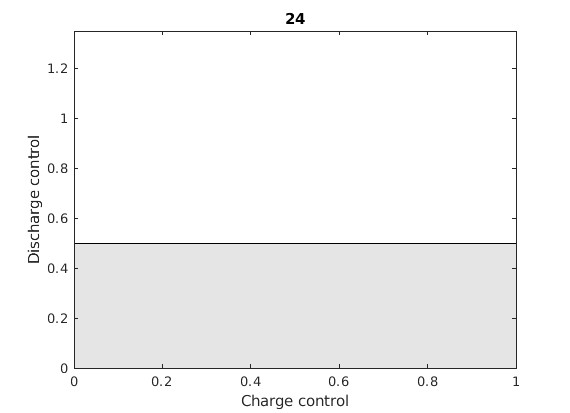}}    \caption{$P = 0$ }
    \label{figDE_1}
    \end{subfigure}
    \hfill
     \begin{subfigure}[b]{0.45\textwidth}
    \centering
    \includegraphics[scale= 0.35]{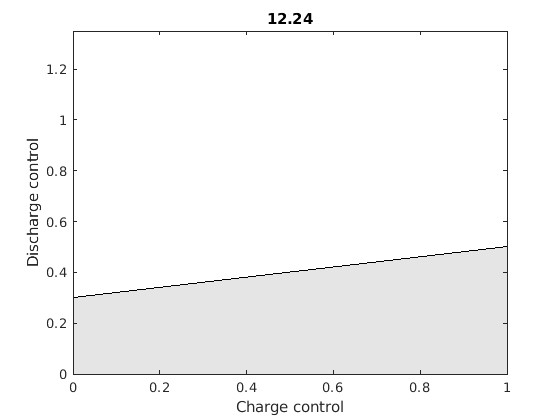}
    \caption{$P > 0$ }
    \label{figDE_2}
    \end{subfigure}
    \caption{Optimization regions}
    \label{figDE}
\end{figure}

{ In general, when  $D \leq E$, we have to find the strategies in the white region such that the infimum is attained. Rearranging the terms in \eqref{inf}, we have to solve $\min_{(a,c) \in \partial^- \mathcal{A}_p, D \leq E} W(a,c)$, with
\begin{gather}
W(a,c) := \left[  a P \left(\eta_c V_s  + e^{-rt} X  \right) - c \left( \frac{V_s}{\eta_d}  + e^{-rt} X \right)  + e^{-rt}( D (X -Z) + X P) \right],
\label{inf_e}
\end{gather}
\noindent while, in the case $D \geq E$, we have to find the minimum in the grey region, i.e. $\min_{(a,c) \in \partial^- \mathcal{A}_p, D \geq E} G(a,c)$, with
\begin{gather}
G(a,c) := \left[  a P \left(\eta_c V_s  + e^{-rt} (X + Z)\right) - c\left( \frac{V_s}{\eta_d} + e^{-rt} (X + Z )\right)  + e^{-rt} (X (D - P) - Z P) \right],
\label{inf_d}
\end{gather}

\noindent to finally find the optimal strategy in the whole control region.


In expressions \eqref{inf_e} and \eqref{inf_d}, we can interpret the terms $ \left(\eta_c V_s  + e^{-rt} X  \right) $ and $\left(\eta_c V_s  + e^{-rt}(X  +Z )\right)$ as the marginal cost associated to store part of the instant power production. 
Instead, the terms $( \frac{V_s}{\eta_d} + e^{-rt}X)$ and $\left( \frac{V_s}{\eta_d} + e^{-rt}(X+Z )\right)$ can be interpreted as the marginal profit of selling the instant stored power. Therefore, according to the sign of these terms we will have different optimal battery management strategies.} {In analyzing these signs, we recall that, since by Lemma 4.1 the value function $V$ is decreasing in $s$, we always have $V_s \leq 0$, while $X$ and $Z$ are also non negative.}

To illustrate how the strategies changes depending on the line \eqref{line}, let us begin by computing the optimal strategies for internal solutions when $P > 0$. 
So, for every fixed $t$, we have:}

{
\begin{itemize}
\item[a.] If $D \leq 0$, i.e. $D = 0$ (recall that $D$ has been modeled to be nonnegative), then $D \leq E$ is verified for all possible values of $a$ and $c$, and we have to solve the problem \eqref{inf_e}, which corresponds to minimizing the linear function $W(a,c)$ in a square. By considering the partial derivatives, we have the minimizers $a^*$ and $c^*$ are at one of the extrema of their respective intervals. More in detail, $a = 0$ if and only if $V_s \geq - \frac{e^{-rt}}{\eta_c} X$, and $c = 0$ if and only if $V_s \geq - e^{-rt} \eta_d X$. The conclusion is that 
$$ (a^*,c^*) = \left\{ \begin{array}{ll}
(0,\Gamma)   & \mbox{ if } - V_s \leq e^{-rt} \eta_d X, \\
(0,0)   & \mbox{ if }  e^{-rt} \eta_d X \leq  - V_s \leq \frac{e^{-rt}}{\eta_c} X, \\
(1,0)   & \mbox{ if } \frac{e^{-rt}}{\eta_c} X \leq - V_s. \\
\end{array} \right. 
$$
Since $- V_s$ is the absolute (recall that $V_s \leq 0$) marginal gain of the value function with respect to the storage level $S$, the interpretation is this: if for each unit of stored electricity the marginal gain $V_s$ is less than the electricity price $X$, net of discounting and of discharge efficiency $\eta_d$, then it is optimal to discharge at the maximum rate $\Gamma$. If instead the marginal gain $V_s$ is more than $X$, net of discounting and of charge efficiency $\eta_c$, then it is optimal to charge at the maximum rate; finally, if both of these thing do not happen, it is better to leave the storage level as it is.

\item[b.] If $0 < D \leq P$, then $D = E$ intersects $\partial^- \mathcal{A}_p$ in $(a,c) = (1 - \frac{D}{P},0)$. When we are in the grey region, i.e. $D \geq E$, we must minimize the linear function $G$, defined in Equation \eqref{inf_d}, on the segment with extrema $(1-\frac{P}{D},0)$ and $(1,0)$: it is thus sufficient to compare
$$ G\left(1-\frac{P}{D},0\right) = (P - D)(\eta_c V_s + e^{-rt} (X + Z) + e^{-rt} (X (D - P) - Z P) $$
with
$$ G(1,0) = P (\eta_c V_s + e^{-rt} (X + Z) + e^{-rt} (X (D - P) - Z P) $$
and it is very easy to see that $G(1,0) \leq G(1-\frac{P}{D},0)$ if and only if $- V_s \geq \frac{e^{-rt}}{\eta_c} (X + Z)$. Instead, when we are in the white region, i.e. $D \leq E$, we must minimize the linear function $W$, defined in Equation \eqref{inf_e}, on the union of the segment with extrema $(0,\Gamma)$ and $(0,0)$ with the segment with extrema $(0,0)$ and $(1-\frac{P}{D},0)$: thus, now we must compare
$$ W(0,\Gamma) = - \Gamma \left(\frac{V_s}{\eta_d} + e^{-rt} X \right) + e^{-rt} (D (X - Z) - X P), $$
$$ W(0,0) = e^{-rt} (D (X - Z) - X P) $$
and
$$ W\left(1-\frac{D}{P},0\right) = (P - D)(\eta_c V_s + e^{-rt} X) + e^{-rt} (D (X - Z) - X P) = G\left(1-\frac{D}{P},0\right). $$
We can see that $W(0,\Gamma) < W(0,0)$ if and only if $\frac{V_s}{\eta_d} + e^{-rt} X > 0$ and $W(1-\frac{D}{P},0) \leq W(0,0)$ if and only if $\eta_c V_s + e^{-rt} X < 0$. Putting all together, we have that the minimum is realized in 
$$ (a^*,c^*) = \left\{ \begin{array}{ll}
(0,\Gamma)   & \mbox{ if } - V_s \leq e^{-rt} \eta_d X, \\
(0,0)   & \mbox{ if }  e^{-rt} \eta_d X \leq  - V_s \leq \frac{e^{-rt}}{\eta_c} X, \\
\left(1 - \frac{D}{P},0\right)   & \mbox{ if }  \frac{e^{-rt}}{\eta_c} X \leq - V_s \leq \frac{e^{-rt}}{\eta_c} (X + Z),\\
(1,0)   & \mbox{ if } \frac{e^{-rt}}{\eta_c} (X + Z) \leq - V_s. \\
\end{array} \right. 
$$
The interpretation is similar to point a., with the addition of the possible optimal control 
$(a^*,c^*) = (1 - \frac{D}{P},0)$, which corresponds to charging the storage with enough intensity to reach the equilibrium $D = E$, which is just enough to obtain the incentive $Z$ for the whole demand $D$; differently from point a., here it is optimal to charge at full rate only when the marginal gain $- V_s$ is greater than $X + Z$. 

\item[c.] If $P \leq D \leq P + \Gamma$, then $D = E$ intersects $\partial^- \mathcal{A}_p$ in $(a,c) = (0,D - P)$. When we are in the grey region, i.e. $D \geq E$, we must minimize the linear function $G$, defined in Equation \eqref{inf_d}, on the union of the segments with extrema $(1,0)$ and $(0,0)$ and the segment with extrema $(0,0)$ and $(0,D - P)$: it is thus sufficient to compare
$$ G(1,0) = P (\eta_c V_s + e^{-rt} (X + Z) + e^{-rt} (X D - (X + Z) P)$$
$$ G(0,0) = e^{-rt} (X D - (X + Z) P) $$
and
$$ G(0,D - P) = - (D - P) \left(\frac{V_s}{\eta_d} + e^{-rt} (X + Z) \right) + e^{-rt} (X D - (X + Z) P)$$
and it is very easy to see that $G(0,0) \leq G(1,0)$ if and only if $- V_s \leq \frac{e^{-rt}}{\eta_c} (X + Z)$ and $G(0,D - P) \leq G(0,0)$ if and only if $- V_s \leq e^{-rt} \eta_d (X + Z)$. Instead, when we are in the white region, i.e. $D \leq E$, we must minimize the linear function $W$, defined in Equation \eqref{inf_e}, on the segment with extrema $(0,\Gamma)$ and $(0,D - P)$: thus, now we must compare
$$ W(0,\Gamma) = - \Gamma \left(\frac{V_s}{\eta_d} + e^{-rt} X \right) + e^{-rt} (D (X - Z) - P X), $$
and
$$ W(0,D - P) = -(D - P) \left(\frac{V_s}{\eta_d} + e^{-rt} X\right) + e^{-rt} (D (X - Z) - P X) = G(0,D - P). $$
We can see that $W(0,\Gamma) \leq W(0,D - P)$ if and only if $- V_s \leq e^{-rt} \eta_d X$. Putting all together, we have that the minimum is realized in 
$$ (a^*,c^*) = \left\{ \begin{array}{ll}
(0,\Gamma)   & \mbox{ if } - V_s \leq e^{-rt} \eta_d X, \\
(0,D - P) & \mbox{ if } e^{-rt} \eta_d X \leq - V_s \leq e^{-rt} \eta_d (X + Z), \\
(0,0)   & \mbox{ if }  e^{-rt} \eta_d (X + Z) \leq  - V_s \leq \frac{e^{-rt}}{\eta_c} (X + Z), \\
(1,0)   & \mbox{ if } \frac{e^{-rt}}{\eta_c} (X + Z) \leq - V_s. \\
\end{array} \right. 
$$
The interpretation is similar to point b., with the difference that the possible optimal control 
$(a^*,c^*) = (0,D - P)$ now corresponds to discharging the storage with enough intensity to reach the equilibrium $D = E$, which is just enough to obtain the incentive $Z$ for the whole demand $D$. 

\item[d.] If $D \geq P + \Gamma$, then $D \geq E$ is verified for all possible values of $a$ and $c$, and we have to solve the problem \eqref{inf_d}. In analogy with case a., this corresponds to minimizing the linear function $G(a,c)$ in a square, so the minimizers $a^*$ and $c^*$ are again at one of the extrema of their respective intervals. More in detail, $a = 0$ if and only if $V_s \geq - \frac{e^{-rt}}{\eta_c} (X + Z)$, and $c = 0$ if and only if $V_s \geq - e^{-rt} \eta_d (X + Z)$. The conclusion is that 
\begin{eqnarray} \label{p_g0_d}
    (a^*,c^*) = \left\{ \begin{array}{ll}
(0,\Gamma)   & \mbox{ if } - V_s \leq e^{-rt} \eta_d (X + Z), \\
(0,0)   & \mbox{ if } - e^{-rt} \eta_d (X + Z) \leq  V_s \leq \frac{e^{-rt}}{\eta_c} (X + Z), \\
(1,0)   & \mbox{ if } \frac{e^{-rt}}{\eta_c} (X + Z) \leq - V_s. \\
\end{array} \right. 
\end{eqnarray}
The interpretation is analogous to point a., with the difference that now each unit of storage has to be evaluated with the sum of electricity price $X$ and incentive $Z$, as any change in the produced electricity $E$ will be fully covered by the demand $D$.
\end{itemize}

{
\begin{remark}
In the above argument, we did not take into account the possible violation of the constraint $a P \leq \alpha_{\mathrm{max}}$. This constraint is typically not violated if the storage facility is large with respect to the photovoltaic plant. If instead the storage facility is underdimensioned with respect to the photovoltaic total capacity, it could be that choices like $a = 1$, or also $a = 1 - \frac{D}{P}$ as in case b., are not admissible. In this case, one should take the minimum between the controls found before and $a^* := \frac{\alpha_{\mathrm{max}}}{P}$ as the admissible optimum. 
\end{remark}
}

The case $P = 0$ can be treated as a limiting case of $P > 0$. The only thing that changes is that, since $a$ is the proportion of photovoltaic production $P$ charging the battery, it will not have any effect in the optimization. For this reason, in the sequel we choose as maximizer $a^* = 0$ to express the fact that we are not
charging the battery.{ Of course, in this case the previous remark will not apply, as we will always have $a P = 0 \leq \alpha_{\mathrm{max}}$. } We can thus summarize the results that we have in this case.
\begin{itemize}
\item[a.] If $D = 0$, then $D \leq E$ is again verified for all possible values of $c$, and we have to solve the problem \eqref{inf_e}, which corresponds to minimizing the linear function $W(a,c)$ in a square, with the difference that now $W$ will be constant in $a$. The conclusion is that 
$$ (a^*,c^*) = \left\{ \begin{array}{ll}
(0,\Gamma)   & \mbox{ if } - V_s \leq e^{-rt} \eta_d X, \\
(0,0)   & \mbox{ if }  e^{-rt} \eta_d X \leq  - V_s \leq \frac{e^{-rt}}{\eta_c} X, \\
\end{array} \right. 
$$

\item[b.] The case $0 < D \leq P$ is now not possible, as $P = 0$.

\item[c.] If $0 \leq D \leq \Gamma$, then $D = E$ intersects $\partial^- \mathcal{A}_p$ in $(a,c) = (0,D - P)$. The analysis proceeds in a similar way than the previous point c., with the final result that the minimum is realized in 
$$ (a^*,c^*) = \left\{ \begin{array}{ll}
(0,\Gamma)   & \mbox{ if } - V_s \leq e^{-rt} \eta_d X, \\
(0,D - P) & \mbox{ if } e^{-rt} \eta_d X \leq - V_s \leq e^{-rt} \eta_d (X + Z), \\
(0,0)   & \mbox{ if }  e^{-rt} \eta_d (X + Z) \leq  - V_s \leq \frac{e^{-rt}}{\eta_c} (X + Z), \\
\end{array} \right. 
$$

\item[d.] If $D \geq \Gamma$, then $D \geq E$ is verified for all possible values of $a$ and $c$, and we have to solve the problem \eqref{inf_d}. The conclusion is that 
\begin{eqnarray} \label{p_g0_d}
    (a^*,c^*) = \left\{ \begin{array}{ll}
(0,\Gamma)   & \mbox{ if } - V_s \leq e^{-rt} \eta_d (X + Z), \\
(0,0)   & \mbox{ if } - e^{-rt} \eta_d (X + Z) \leq  V_s \leq \frac{e^{-rt}}{\eta_c} (X + Z), \\
\end{array} \right. 
\end{eqnarray}
\end{itemize}
}

\subsection{Numerical implementation}

We apply Kushner's scheme to obtain a numeric approximation of the value function $V$ in \eqref{V}. This scheme approximate the controlled Markov diffusion by a controlled Markov chain on a lattice in $\mathbb{R}^n$. It turns out that under suitable assumption the dynamic programming equation associated to the Markov chain, which leads in the finite difference scheme, is related with the HJB equation of the original problem in continuous time (see \cite{kdnm} for details). To prove convergence of the approximate solution to the value function, in \cite{kdnm} they use a probabilistic approach, while in \cite{FS} they apply viscosity solution techniques.

The convergence of the scheme is guaranteed thanks to \cite[Theorem 4.1, Chapter IX]{FS}. To apply it to our case it is needed to consider the time as another state variable, which is a natural choice in implicit Markov chain approximations. The interested reader is refereed to   \cite{kdnm, FS} for details.


Now we briefly describe the scheme approximation for the HJB equation \eqref{HJB}. Let us consider a time step $\tau > 0$ and set $t_i = t + \tau (i - 1)$, with $i = {1 , \ldots, N_t}$ and $\tau = (T - t)/ N_t $. Consider the spacial step $\xi$ for the photovoltaic production and $\delta$ for the state of charge.  Set $p_n = \xi (n -1 )$, with $n = {1, \ldots, N_p}$ and $\xi = Y/ N_p $; and $s_k = S_{min} + \delta (k - 1)$, with $k = {1, \ldots, N_s}$ and $\delta = \frac{(S_{max} - S_{min})}{N_s}$.
 
Let
\begin{eqnarray*}
    f^+ = \max \{ f, 0  \} \mbox{ and } f^- = \max \{ -f, 0  \}, 
\end{eqnarray*}

\noindent be the positive and negative part of function $f$.  Let us consider for any function $U(t,p,s)$,

\begin{eqnarray*}
\Delta_t^-U & = & \frac{U(t,p,s) - U_(t-\tau,p,s)}{\tau},\\
\Delta_p^+U  & = & \frac{U(t-\tau,p + \xi,s) - U(t-\tau,\xi,s)}{\xi},\\
\Delta_s^+U  & = & \frac{U(t-\tau,p ,s+\delta) - U(t-\tau,\xi,s)}{\xi},\\
\Delta_s^-U  & = & \frac{U(t-\tau,p ,s) - U(t-\tau,\xi,s-\delta)}{\xi},\\
\Delta_{pp}U  & = & \frac{U(t,p + \xi,s) - 2U(t,p,s) + U(t,p-\xi,s)}{\xi^2}.
\end{eqnarray*}

\noindent Consider as final condition

\begin{eqnarray*}
 U(N_t,p_n,s_k) := 0 \mbox{ , for } n = 1, \ldots, N_p \mbox{ and } k = 1, \ldots, N_s.
\end{eqnarray*}

\noindent Moreover, we consider the approximation for the second derivatives in $p$ in the border when $n = 1$, 

\begin{eqnarray*}
\Delta_{pp}U & = & \frac{U(t - \tau,p + 2 \xi,s) - 2U(t - \tau,p +\xi,  s) + U(t - \tau, p, s)}{\xi^2}
\end{eqnarray*}

\noindent and the approximation for the derivatives in the border when $n = N_p$, 

\begin{eqnarray*}
\Delta_{pp}U & = & \frac{U(t - \tau,p,s) - 2U(t - \tau,p -\xi,  s) + U(t - \tau, p - 2\xi, s)}{\xi^2}.
\end{eqnarray*}

\noindent In addition, we consider the first order derivative in $p$, when $n = N_p$, as

\begin{eqnarray*}
    \Delta_p^-U  & = & \frac{U(t-\tau,p ,s) - U(t-\tau,p - \xi,s)}{\xi}
\end{eqnarray*}
 
\noindent Set $f_s =  \eta_c a f_p(t_i)e^{p_n} - \frac{c}{\eta_d}$. We obtain the following semi implicit finite difference scheme for $i = \{ 2, \ldots\, T\}$, $n = \{ 2, \ldots\, N_p -1\}$, $k = \{ 1, \ldots\, N_s - 1\}$,

\begin{eqnarray}
& & \frac{U(t_{i-1},p_n,s_k)}{\tau} + \xi_p p_n \left(  \frac{ U(t_{i-1},p_{n+1},s_k) - U(t_{i-1},p_n,s_k) }{\xi} \right) \nonumber\\
& & - \frac{\sigma_p^2}{2} \left(  \frac{ U(t_{i-1},p_{n+1},s_k) - 2 U(t_{i-1},p_n,s_k) +  U(t_{i-1},p_{n-1},s_k)}{\xi^2} \right)\nonumber \\
& & = \frac{U(t_i,p_n,s_k)}{\tau} + \inf_{A_p} \left[ f_{s}^+\left(\frac{U(t_i,p_n,s_{k+1}) - U(t_i,p_n,s_k)}{\delta}\right) - f_{s}^-\left(\frac{U(t_i,p_n,s_k) - U(t_i,p_n,s_{k-1})}{\delta}\right) \right. \nonumber \\
& & \left. + \hat{h}_1(t_i,p_n,a,c) -Z  \hat{h}_2(t_i,p_n,a,c)\right].
\label{semi_imp}
\end{eqnarray}

The approximation in the border $n = \{ 1, N_p\}$ and $k = N_s$ is as in \eqref{semi_imp}, but with the respective derivative approximations.

\section{Application} \label{app}

In this section we present the results of the numerical implementation using realistic data, in the sense that the parameters of the equation describing the evolution of the electricity price, power demand and photovoltaic production where estimated using real observations. For the sinusoidal deterministic function $f_x$, $f_d$ and $f_p$, we use harmonic regression and maximum likelihood estimators \cite{ts}. For the estimation of the exponential Ornstein-Ulhenbeck we use methods described in \cite{brigo}, which produce maximum likelihood estimators.

Instead, we consider the efficiency parameter of a ion-lithium battery which is normally used in photovoltaic systems since its performance is more effective and efficient with respect to other storage technologies \cite{ion_li_batt}.

\subsection{Data description and parameter estimations}

\paragraph{The electricity price.} To estimate the seasonal component of the electricity spot price \eqref{Xs} we use hourly measurements of the Italian unique national price (PUN) measured in \euro/MWh, from 1st January 2023 to 31st January 2023. The data is available at \cite{GME}. As we are interested in the daily seasonalities of the electricity price, we only consider frequencies inside this time lapse, so yearly effects as summer or winter periods (see Figure \ref{ED_season}) were not taken into account. As expressed in Equation \eqref{Xs}, for the seasonal component of the electricity spot price, we fit the following deterministic curve to the data

\begin{eqnarray*}
    f_x(u) = e^{\sum_{i = 1}^3 \alpha_i \sin(2 \pi \psi_i u)+ \beta_i \cos(2 \pi \psi_i u )}.
\end{eqnarray*}

\noindent for $i = \{ 1, 2, 3\}$, where $\psi_i$ are daily frequencies, $\alpha_i$ and $\beta_i$ the sin and cosine amplitudes respectively. 

\begin{table}[h]
    \centering
    \begin{tabular}{|c|c|c|}
    \hline
        Frequency $\psi$ & Amplitude $\alpha$ & Amplitude $\beta$  \\
        \hline
         0.04167 & $-$0.30068  & $-$0.09365\\
         \hline
         0.08333 & $-$0.21155 & 0.09567\\
         \hline
         0.125 &  0.07929 & 0.07220\\
         \hline     
    \end{tabular}
    \caption{Estimated parameters for the seasonal part of the electricity price}
    \label{season_parameters_ep}
\end{table}

\begin{figure}[h]
    \centering
    \includegraphics[scale = 0.4]{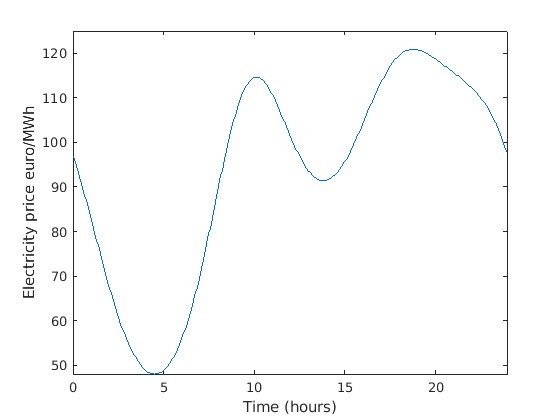}
    \caption{Estimated electricity price daily profile}
    \label{fig_ep}
\end{figure}

\begin{figure}[h]
    \centering
    \includegraphics[scale = 0.4]{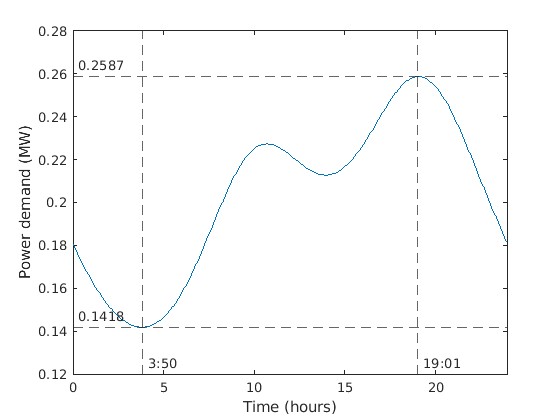}
    \caption{Estimated power demand daily profile}
    \label{fig_d}
\end{figure}

Table \ref{season_parameters_ep} summarized the estimated parameters. The significant frequencies correspond to periods of 24, 12, and 8 hours, respectively. Figure \ref{fig_ep} shows the daily profile obtained with the estimated values. We observe two price peaks: one around midday and the second during the evening. These peaks are correlated with the residential power demand profile: as the power demand increases, increases also the electricity price. Compare Figures \ref{fig_ep} and \ref{fig_d}.

\paragraph{Power demand.} To estimate the seasonal component of the power demand we consider hourly measurements of the power demand in Italy, measured in MW, from 1st January to 31st January 2023. The data is available at \cite{GME}. As in the electricity price model, we are interested in the daily seasonalities, monthly or yearly seasonalities were not considered. According to Equation\eqref{Ds}, the seasonal part of the power demand is described by a deterministic function:
\begin{eqnarray*}
    f_d(u) = e^{  \sum_{i = 1}^3 \alpha_i \sin(2 \pi \psi_i u)+ \beta_i \cos(2 \pi \psi_i u )}
\end{eqnarray*}

\noindent for $i = \{ 1, 2, 3\}$, where $\psi_i$ are daily frequencies, $\alpha_i$ and $\beta_i$ the sin and cosine amplitudes, respectively. 

The results of the parameters estimation are presented in Table \ref{season_parameters_d}. We found three main daily significant frequencies: 24, 12 and 8 hours, respectively.

\begin{table}[h]
    \centering
    \begin{tabular}{|c|c|c|}
    \hline
        Frequency $\psi$ & Amplitude $\alpha$ & Amplitude $\beta$  \\
        \hline
         0.04167 & $-$0.21109 & $-$0.10399 \\
         \hline
         0.08333 & $-$0.12501 & 0\\
         \hline
         0.125 &  0.02541  & 0\\
         \hline
    \end{tabular}
    \caption{Estimated parameters for the seasonal part}
    \label{season_parameters_d}
\end{table}

Figure \ref{fig_d} shows the function obtained by means of the harmonic fit, which exhibit a common behavior in residential consumption. We observe the minimum consumption of $0.1418$ MW during the night, around $3:50$ AM ($0.16$ days); then a morning peak of $0.2273$ MW around $10:40$ AM ($0.445$ days), related with the usual activities of people waking up in the morning, followed by a decrement in the consumption, since in residential building a great part of the residents is out for working or other activities. From the afternoon, around $13:55$ ($0.58$ days), the power consumption increases until it reaches its daily maximum of $0.2587$ MW in the evening at $19:01$ (0.7925 days). For residential consumption profile, this maximum is related to the moment when most of the people come back home. As we use data corresponding to a winter period we observe a winter power consumption profile. 

\paragraph{Photovoltaic production.} To estimate the seasonal part of the photovoltaic production and the parameters of the exponential Ornstein- Uhlenbeck process, we consider simulated photovoltaic power production measurements every 5 min from 1st January to 31st January 2023, from a location in Florida, USA available at \cite{Data_photo}. We consider just one significant frequency representing the period of $f = 1/24$ hours.

\begin{eqnarray} \label{photo_eq}
    P(u) = A \sin( 2 \pi \psi (u + \phi)) e^{U_p(u)}
\end{eqnarray}

\begin{table}[h]
    \centering
    \begin{tabular}{|c|c| c | c | }
    \hline
        Parameter & Value & Parameter    & Estimated value  \\
        \hline
         A & 0.5 (MW) &  $\xi_p$  & 2 \\
         \hline
         $\phi$ & 18 (h) & $\sigma_p$ & 0.3\\
         \hline
    \end{tabular}
    \caption{Estimated parameters for the photovoltaic production}
    \label{sppoup}
\end{table}

Table \ref{sppoup} summarized the value of the parameters estimated for the photovoltaic production. The drift $\xi_p$ and volatility $\sigma_p$ for the Ornstein-Uhlenbeck process are in a daily scale.

\begin{figure}[H]
    \centering
    \includegraphics[scale = 0.4]{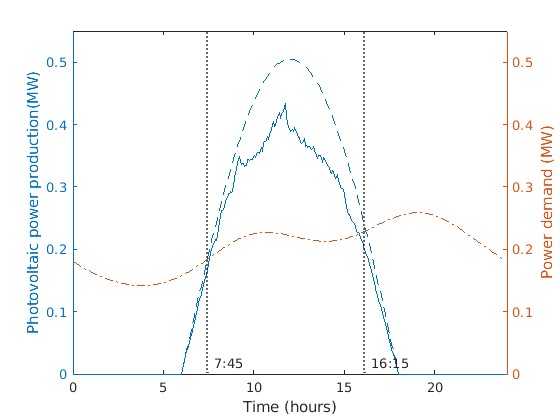}
    \caption{Self-power production and power consumption}
    \label{photo_sim}
\end{figure}

In Figure \ref{photo_sim} we present the expected value of the photovoltaic production  for $p = 0$ \eqref{EPp} (dashed  blue curve), while in solid line it is presented one simulated trajectory of the process describing the photovoltaic production \eqref{photo_eq}. The dashed red line represents the power demand of the group. The vertical dashed line at time 7:45, represents the time during the day when the expected photovoltaic production exceed the power demand, while the dashed vertical line at 16:15 is the time when the expected photovoltaic production does not cover the power demand anymore. We highlight this times in order to analyze, in the next section, how changes the battery management strategy during different hours of the day.  

\paragraph{Battery parameters}

We consider a lithium-ion battery, which is widely used in photovoltaic systems applications. We consider a  battery as \cite{ion} and we take as maximum charging power $\alpha_{max}$, maximum discharging power $\Gamma$ and energy capacity $S_{max}$ rounded values from the technical specifications. Instead, the efficiency parameters where considered following \cite{eta_cd}. Table \ref{tab_bat} summarized all the technical specification for the battery.

\begin{table}[h]
\centering
\begin{tabular}{| c | c |}
\hline
Parameter & Value \\
\hline
$\eta_c$ & $99 \%$ \\
\hline
$\eta_d$ & $97 \% $\\
\hline
$\alpha_{max}$ & $0.01$ MW  \\
\hline
$\Gamma$ &  $0.028$ MW \\
\hline
$S_{max}$ & $0.03$ MWh\\
\hline 
$S_{min}$ & 0 MWh\\
\hline
\end{tabular}
\caption{Battery parameters}
\label{tab_bat}
\end{table}

\paragraph{Photovoltaic system} We suppose a photovoltaic system composed by a photovoltaic capacity with a maximum steady state average power production of $0.5057$ MW and two batteries connected in parallel, with the specification as in Table \ref{tab_bat}. As our application is set in the Italian context we consider a profitable sizing according to reported studies as \cite{Lage}.

\begin{remark} \label{remark_domanda} In our application for night periods, i.e., $ u \in [0,7]$ such that $P(u) = 0$, we have $\min_{u \in [0,7]_ {s.t. P(u) = 0}} D(u) > \Gamma$, since the minimum power demand attained at $03:43$ is $0.1418$ MW and the maximum discharging power of two batteries connected in parallel with the specification in Table \ref{tab_bat} is $0.084$ MW. Hence, for internal solutions, the optimal discharging action during the night will be either discharge the battery at its maximum power or do not discharge it at all (see Section \ref{small}).
\end{remark}

\subsection{Results} \label{results}

\begin{figure}[H]
    \centering
    \includegraphics[width=0.5\linewidth]{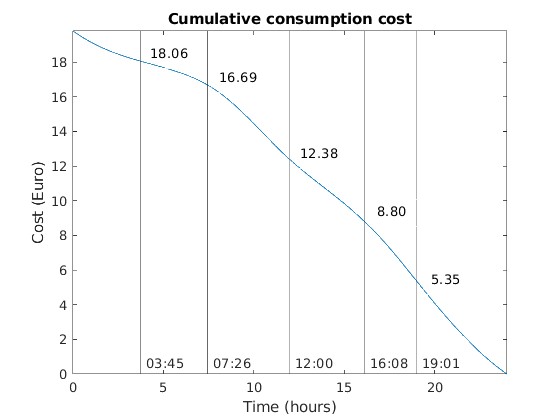}
    \caption{Cumulative cost without photovoltaic system and critical hours}
    \label{cost}
\end{figure}

In this section we present and discuss the solution obtained by implementing computationally the scheme \eqref{semi_imp}. We consider a time interval of one day, starting from $t = 0$ to $T = 1$, and all the parameters values presented in the section before (Tables \ref{season_parameters_ep}, \ref{season_parameters_d}, \ref{sppoup}, \ref{tab_bat}). We take as time step $\tau = 0.001$ and as state space steps $\xi = 0.04$ and $\delta = 0.005$, for the photovoltaic production and the state of charge, respectively. Since we only consider daily seasonalities we obtain the same strategies for the seven days of the week. Therefore, we present the results in one day for different hours. We will take into account the following times during the days: minimum power demand (03:43), first instant in the day that the average photovoltaic production, with starting point $p = 0$, exceed the power demand (7:26); maximum expected power production (12:00), first instant during the evening such that the photovoltaic production does not cover the power demand  (16:08) and the maximum power demand peak, attained at 19:01. In order to assess the reduction in consumption cost by optimally manage the battery in the photovoltaic system, in Figure \ref{cost} are presented the cumulative cost during the first day without a photovoltaic system. The vertical lines represents the critical instant in the day that we are going to discuss, with the respective cumulative cost at that moment. Recall that the closer we are to the final time $T$, the lowest will be the cost value, since the problem is solve backward.

\subsubsection{Minimum power demand}  
\begin{figure}[H]
    \centering
    \includegraphics[width=0.5\linewidth]{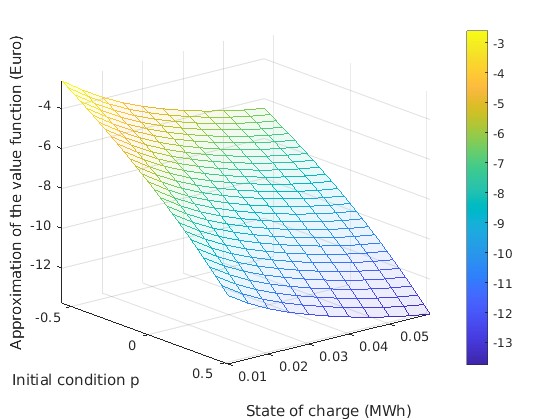}
    \caption{Approximated value function at minimum power demand}
    \label{V_03}
\end{figure}

Here we analyze the optimal management strategy when the power demand reaches its daily minimum. In the simulation this point is attained at $u =0.155 $, corresponding to the time 03:45. Therefore, in this case the only admissible charging action is $a^* = 0$. Compared with the cumulative cost during the first day at this time $ 18.06$\euro, we obtain a total cover of the cumulative consumption cost, plus profits ranging from a minimum of $1.5568$ \euro, for fully discharge battery and $p =  -0.56$ and a maximum of $13.7515$\euro, for a battery fully charged and $p = 0.52$ (see Figure \ref{V_03} showing the approximated value function). 

\begin{figure}[H]
\centering
     \begin{subfigure}[b]{0.45\textwidth}
         \centering
         \includegraphics[width=\textwidth]{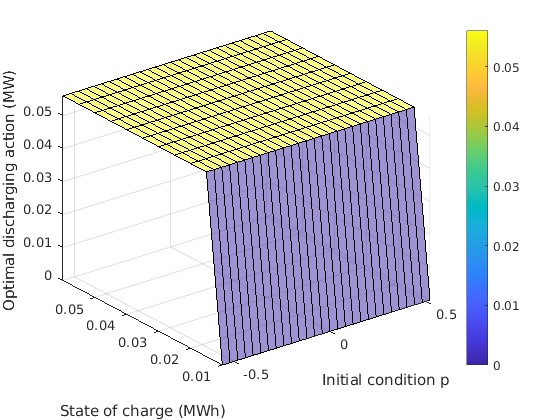}
         \caption{Optimal management strategy at the minimum power demand}
         \label{mpd_1}
     \end{subfigure}
    \hfill
    \begin{subfigure}[b]{0.45\textwidth}
         \centering
         \includegraphics[width=\textwidth]{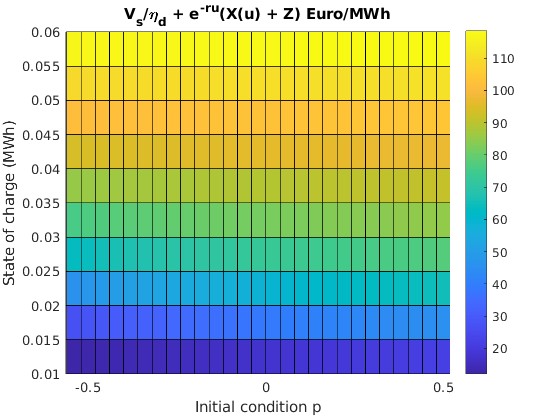}
         \caption{Marginal profit with incentive of selling the stored energy }
         \label{mpd_f}
    \end{subfigure}
\caption{Discharging action and surface of marginal profit with incentive}
\label{mpd_dis}
\end{figure}

Figures \ref{mpd_1} shows the discharge action. We observe an "all or nothing" discharge action, as it was expected from Remark \ref{remark_domanda}. According to Section \ref{small}, when the power demand is higher than the available power in the battery, we will discharge it at its maximum rate if the marginal profits considering incentives is positive, i.e., when  $ \frac{V_s}{\eta_d} + e^{-rt} (X(t) + Z) > 0$. In Figure \ref{mpd_f} we observe that the marginal profit is positives and therefore the optimal strategy is to discharge the batteries at the maximum rate of 0.056 MW. Of course when the batteries are fully discharge the only admissible strategy during the night is to do not perform any action. 

\subsubsection{Photovoltaic production exceed power demand} 
\begin{figure}[H]
    \centering
    \includegraphics[width=0.5\linewidth]{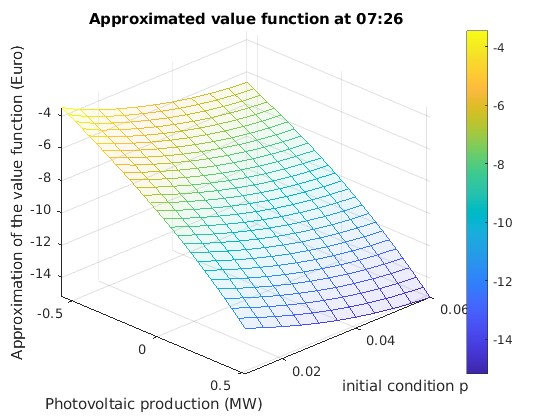}
   \caption{Approximated value function at when photovoltaic production exceed power demand}
    \label{V_04}
\end{figure}

Here we analyze the optimal battery management when the expected value of the photovoltaic production with $p = 0$, exceed the power demand, i.e., the first moment such that $\mathbb{E} [P(u)] > D(u)$. In the simulation this point is attained at $u =0.31$, corresponding to the time $07:26$. Figure \ref{V_04} shows the approximated value function at the analyzed time. We observe that the photovoltaic system produce profits ranging from \euro$2.13$   to \euro$15.19$,  depending on the energy stored in the battery. As in the previous case, the energy sold to the grid is enough to cover the power consumption cost and produce profits from the sales during the day.
 
\begin{figure}[H]
\centering
     \begin{subfigure}[b]{0.45\textwidth}
         \centering
         \includegraphics[width=\textwidth]{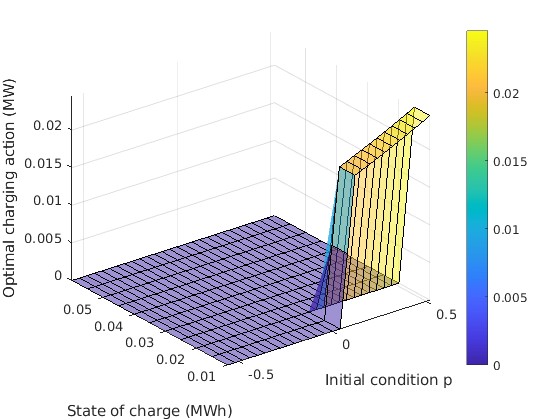}
         \caption{Charging action}
         \label{a_s_1}
     \end{subfigure}
    \begin{subfigure}[b]{0.45\textwidth}
         \centering
         \includegraphics[width=\textwidth]{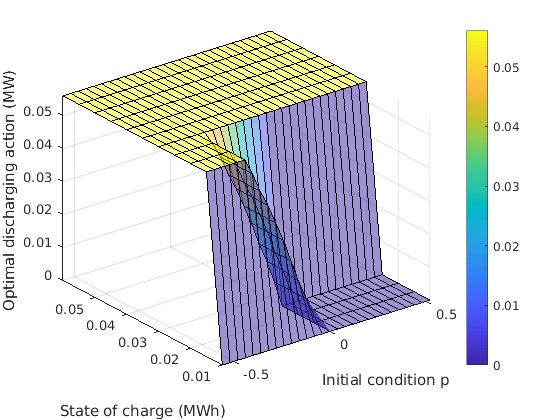}
         \caption{Discharging action}
         \label{c_s_1}
    \end{subfigure}
    \caption{Optimal battery management}
\label{ch_dis_13}
\end{figure}

Regarding the optimal battery management, Figure \ref{ch_dis_13} shows the obtained results. 

\begin{figure}[H]
\centering
     \begin{subfigure}[b]{0.45\textwidth}
         \centering
         \includegraphics[width=\textwidth]{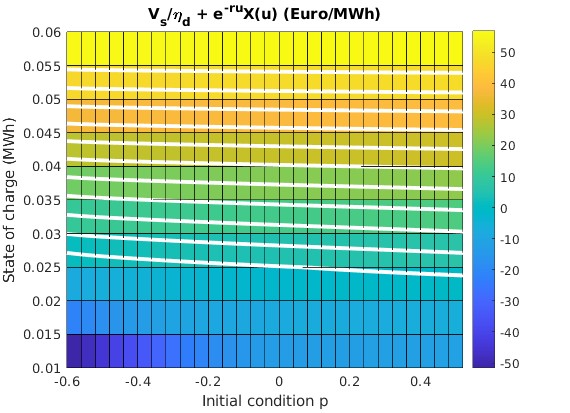}
         \caption{Marginal profit without incentives}
         \label{cond_2_0726}
     \end{subfigure}
    \begin{subfigure}[b]{0.45\textwidth}
         \centering
         \includegraphics[width=\textwidth]{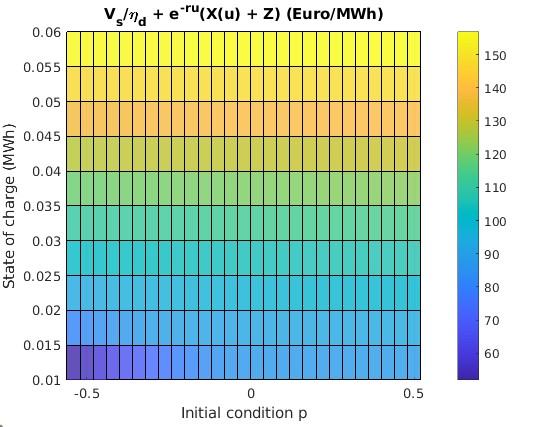}
         \caption{Marginal profit with incentives}
         \label{c_2_s_0726}
    \end{subfigure}
    \caption{Marginal cost and profits}
\label{cp_0726}
\end{figure}

In Figure \ref{ch_dis_13} we observe three cases: at $p = 0$ the power demand matches the average photovoltaic production, while for $p > 0$ the power demand is lower than the produced energy and for $p < 0$, the power demand is higher than the average production.  According to Section \ref{small}, in the case $p \geq 0$, when the marginal costs without incentives are negatives and the marginal costs with incentives are positive, it is optimal to charge enough energy in such a way that the energy which is sold to the grid equals the power demand ($D(u) = E(u)$). On the other hand, it is optimal to fully discharge the battery if both marginal profits are positive. In Figure \ref{cond_2_0726}, the white level curves represent positive values for the marginal profit without incentives.

\subsubsection{Maximum power production}  
\begin{figure}[H]
    \centering
    \includegraphics[width=0.5\linewidth]{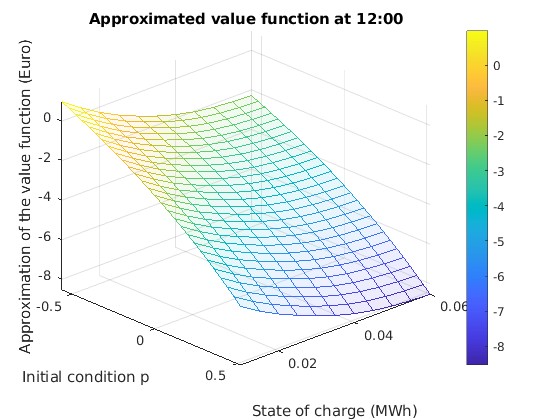}
    \caption{Approximated value function at midday}
    \label{U_1200}
\end{figure}

Here we analyze the optimal management of the battery when the photovoltaic production reach its maximum expected value. In the simulation this point is attained at $t =0.5$, corresponding to midday. In Figure \ref{U_1200}, showing the approximated value function, we observe a maximum profit of \euro $8.5325$ for fully charge batteries and initial condition $p = 0.5$. However, the photovoltaic system is not enough to cover the cumulate consumption cost in one day: we observe the maximum cost equal to \euro $2.46$, for fully discharged batteries and initial condition $p = -0.5$.

\begin{figure}[H]
\centering
     \begin{subfigure}[b]{0.45\textwidth}
         \centering
         \includegraphics[width=\textwidth]{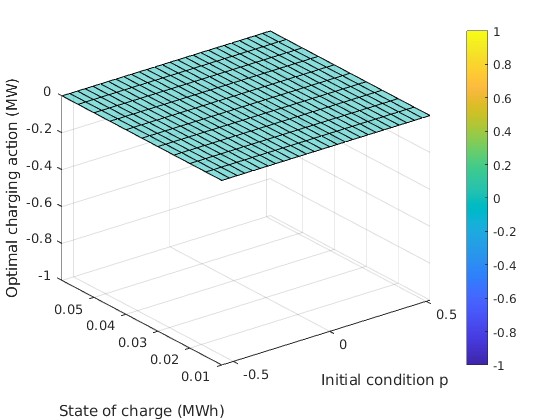}
         \caption{Charging action}
         \label{a_mp_12}
     \end{subfigure}
     \hfill
    \begin{subfigure}[b]{0.45\textwidth}
         \centering
         \includegraphics[width=\textwidth]{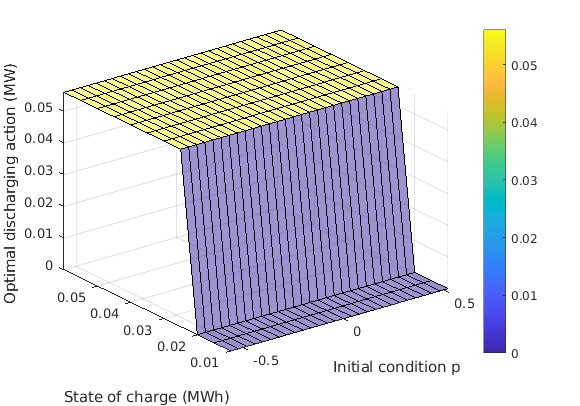}
         \caption{Discharging action}
         \label{c_mp_12}
    \end{subfigure}
\caption{Optimal battery management}
\label{ac_mp_12}
\end{figure}

\begin{figure}[H]
\centering
         \includegraphics[width=0.5\linewidth]{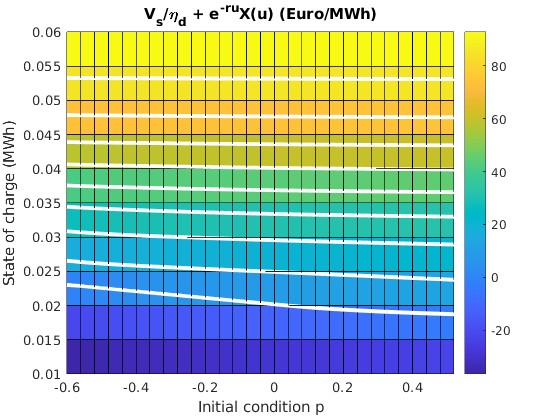}
         \caption{Marginal profit without incentives}
         \label{cond_2_12}
\end{figure}

Figure \ref{ac_mp_12} shows the charging and discharging optimal actions. From Figure \ref{a_mp_12} we observe that it is not optimal to charge the battery, instead it is more convenient to sell all the photovoltaic production to the grid. On the other hand, from Figure \ref{c_mp_12}, we observe that it is optimal to discharge the battery for values of the state of charge grater than $0.02$ MWh, while for lower values it is not optimal to discharge the battery. Let us compare this results with the one obtained in Section \ref{small}. At this time it is verified that $P(u) > D(u)$, therefore we expect that for all those values where it is optimal to discharge the battery we will have positive profits without incentives, while in the region where it is optimal to do not perform any action, we expect to have negative profits and positive cost without incentives. Figure \ref{cond_2_12} shows the profit in color scale. The region where it is optimal to discharge coincide with the region where the marginal profits are positive,i.e., $\left( \frac{V_s}{\eta_d} + e^{-rt}X(t) \right) > 0$. The white lines are positive curve levels.


\subsubsection{Demand exceed photovoltaic production}
\begin{figure}[H]
    \centering
    \includegraphics[width=0.5\linewidth]{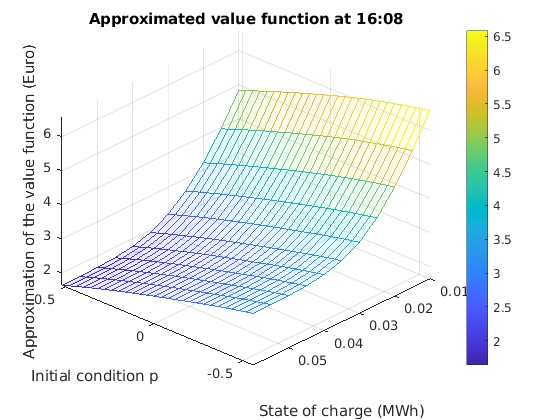}
   \caption{Approximated value function at when power demand exceed  photovoltaic production }
    \label{V_05}
\end{figure}

Here we analyze the optimal battery management when the expected value of the photovoltaic production with $p = 0$, goes down the power demand, i.e., the first moment in the evening such that $P(u) < D(u)$. In the simulation this point is attained at $u =0.6725$, corresponding to the time $16:08$. Figure \ref{V_05} shows the approximated value function at the analyzed time. We observe that the photovoltaic system is able to reduce the consumption cost up to \euro $1.64$ for $p = 0.5$ and fully charged batteries. The maximum cost of \euro$7.45$ is for $p = 0.5$ and fully discharged batteries. We obtain just one euro of saving compared with the cumulative cost without a photovoltaic system.

\begin{figure}[H]
\centering
     \begin{subfigure}[b]{0.45\textwidth}
         \centering
         \includegraphics[width=\textwidth]{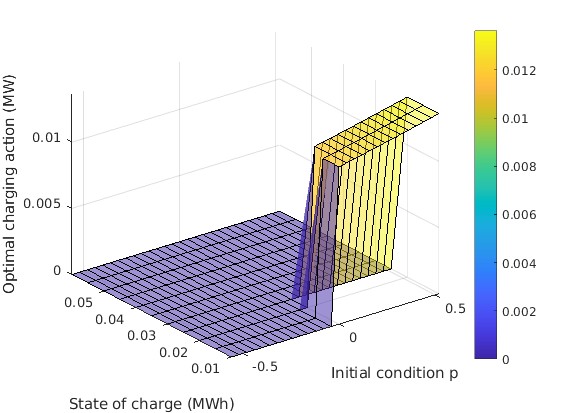}
         \caption{Charging action}
         \label{a_s_16}
     \end{subfigure}
    \begin{subfigure}[b]{0.45\textwidth}
         \centering
         \includegraphics[width=\textwidth]{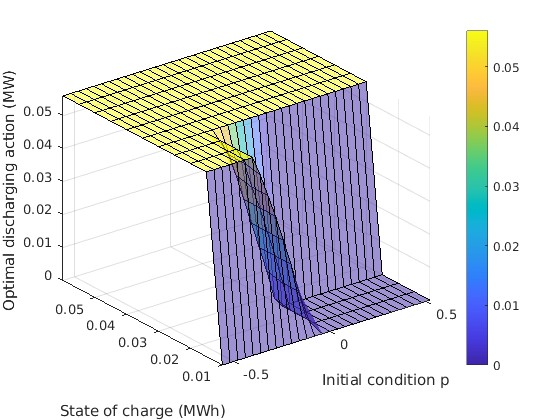}
         \caption{Discharging action}
         \label{c_s_16}
    \end{subfigure}
    \caption{Optimal battery management}
\label{ch_dis_16}
\end{figure}

Regarding the optimal battery management, Figure \ref{ch_dis_16} shows the obtained results. For $p > 0$ the average photovoltaic production is still enough to cover the power demand. Instead for $p \leq 0$ the power demand is grater than the produced energy.

\begin{figure}[H]
\centering
     \begin{subfigure}[b]{0.45\textwidth}
         \centering
         \includegraphics[width=\textwidth]{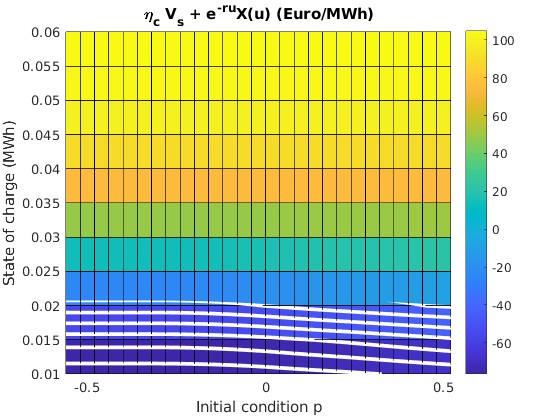}
         \caption{Marginal cost without incentives}
         \label{c_1_1608}
     \end{subfigure}
    \begin{subfigure}[b]{0.45\textwidth}
         \centering
         \includegraphics[width=\textwidth]{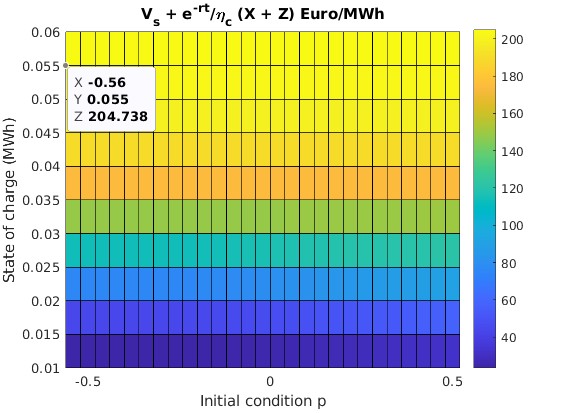}
         \caption{Marginal cost with incentives}
         \label{c_1_s_1608}
             \end{subfigure}
    \caption{Marginal costs of storing energy}
\label{cp_1608}
\end{figure}

\begin{figure}[H]
\centering

     \includegraphics[scale=0.45]{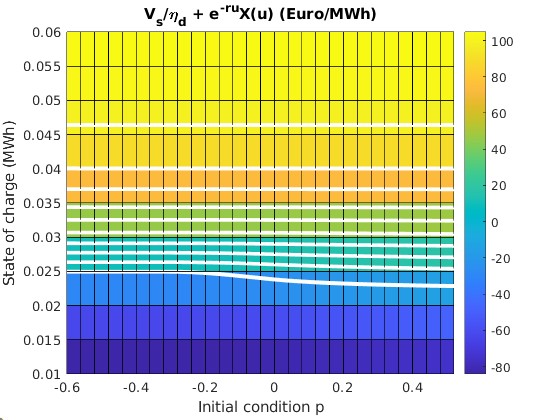}
         \caption{Marginal profits without incentives}

\label{cond_2_1608}
\end{figure}

We are going to comment the result for $p > 0$. According to Section \ref{small}, when the marginal cost are negative, but positive considering the incentives, then it is optimal to charge just enough energy in order to match he power demand with the sold energy, i.e., $D(u) = E((u)$. This is verified by observing Figure \ref{cp_1608}. On the other hand, when the profit without incentive are positive, then it is optimal to discharge at maximum rate. Observe in Figure \ref{cond_2_1608} that this profits are positive from $s = 0.23$ and all $p  > 0$. The white level curves correspond to positive values for this surface.


\subsubsection{Evening peak of the power demand} 
\begin{figure}[H]
    \centering
    \includegraphics[width=0.5\linewidth]{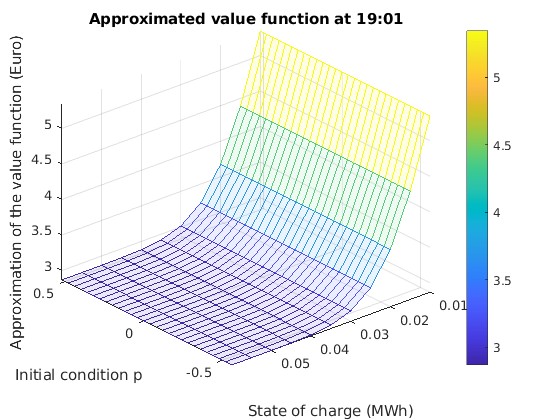}
    \caption{Approximated value function at the power demand peak}
    \label{U_1902}
\end{figure}

Here we analyze the  optimal management action when the power demand reaches its daily maximum. In the simulation this point is attained at $t =0.7925$, corresponding to the time 19:01. At this time without a photovoltaic system the cumulative consumption cost of the group is \euro$5.35$.  Figure \ref{U_1902}
shows the cumulative cost considering the photovoltaic system. We observe that we obtain reduced cost of \euro$2.87$ by optimal discharging the battery when it is fully charged. Instead, when the battery os fully discharged we do not are able to cover the consumption cost and we obtain the same as in the case without the photovoltaic system. Therefore, in the best scenario, we are able to cover at least $53 \%$ of the consumption cost.

\begin{figure}[H]
\centering
     \begin{subfigure}[b]{0.45\textwidth}
         \centering
         \includegraphics[width=\textwidth]{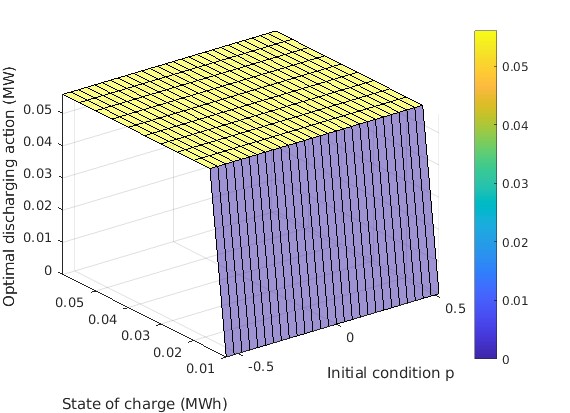}
\caption{Discharging action}
         \label{dis_epdp}
     \end{subfigure}
     \hfill
    \begin{subfigure}[b]{0.45\textwidth}
         \centering
         \includegraphics[width=\textwidth]{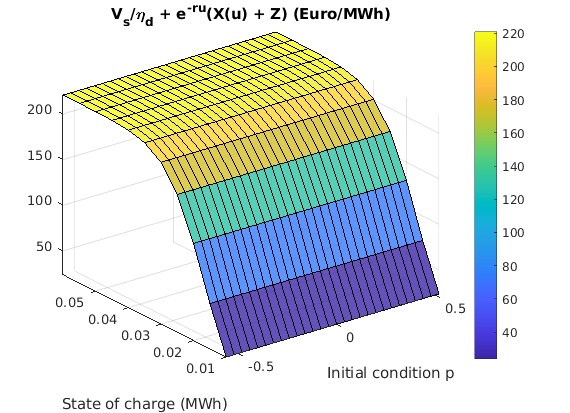}
    \caption{Marginal profits of selling the stored energy}
         \label{conds_peve}
    \end{subfigure}
\caption{Discharging action and marginal profits with incentives}
\label{c_and_cond_eve}
\end{figure}

According to our model at this time there is no more sunlight, therefore the only admissible charging action is $a^* = 0$. Figure \ref{dis_epdp} shows the discharging actions during the evening peak of the power demand. 

As the power demand peak is equal to to $D(u) = 0.2587$ MW, but the maximum discharging power of the batteries is $0.0560$ MW, we have $D(u) > \Gamma$. According to Section \ref{small}, when the marginal profit of selling the stored energy with incentives is positive, it is optimal to discharge the battery at its maximum power. We can verify this by observing Figure \ref{conds_peve}. We observe that the marginal profit of selling the stored energy considering the incentive is always positive, i.e., the condition $ \frac{V_s}{\eta_d} + e^{-rt}(X(t) + Z) >0 $ for maximum rate discharge action is respected.

\section{Conclusions} \label{conclusion}

In this work, we study the optimal management of the battery presented in a photovoltaic system. We model the problem as a finite horizon stochastic optimal control problem, where we model the stochastic behavior of the photovoltaic production as a mean reverting process. We were able to establish the condition for marginal profits and costs such that it is optimal to charge or discharge the battery. The numerical results validate the theoretical ones: the battery charging and discharging management follows the marginal profit and cost conditions presented in Section \ref{small}. Additionally, we obtain optimal controls of the form: either we charge the battery or we discharge the battery, but both actions are not performed simultaneously since it is not optimal. Moreover, the monotonicity conditions of the value function in $p$ and $s$ are also respected. The results show that the cumulative costs are decreasing in $p$ and are convex and decreasing in $s$. 

From a technical point of view, we conclude that it is optimal to maintain the charge of the battery close to the minimum allowed level. Therefore, when it is optimal to perform the charging-discharging action we found that we charge the battery only when the energy level is close to $S_{min}$, while for higher energy levels, we discharge the battery.

From an economic point of view, we conclude that the optimal management of the photovoltaic system considered in this work, covers consumption costs and, at certain times in the day, it also generate revenues. 

As future work, we propose the inclusion of the stochastic behavior of the electricity price and power demand, as well as a more detailed model of the energy level in the battery, considering the wear suffer by the battery due to charging and discharging actions and the natural degradation of this device.

\section*{Acknowledgements}

The second author is supported by the Italian Minister of
University and Research (MUR) through the following projects: "A geo-localized data framework for managing climate risks and designing policies to support sustainable investments" (No. 20229CWYXC) within the PRIN 2022 program; "Fin4Green - Finance for a Sustainable, Green and Resilient Society Quantitative approaches for a robust assessment and management of risks related to sustainable investing", (No. 2020B2AKFW-003)
within the PRIN 2020 program. Her research was also supported by the European COST (Cooperation in Science and Technology through the project "Fintech and Artificial Intelligence in Finance –Towards transparent financial Industry", within the COST Action CA19130. 
The third author 
is supported by the INdAM - GNAMPA Project code CUP E53C23001670001 and by the projects funded by the European Union - Next Generation EU, Mission 4 Component 1, 2022BEMMLZ ``Stochastic control and games and the role of information'' CUP C53D23002430006 and P20224TM7Z ”Probabilistic methods for energy transition”,  CUP C53D23008390001. 
The authors also wish to thank  Michele Aleandri,  Bernardo D'Auria and Paolo Falbo for fruitful discussions.

\end{document}